\documentclass[fleqn,usenatbib]{mnras}
\usepackage{newtxtext,newtxmath}
\usepackage[T1]{fontenc}
\DeclareRobustCommand{\VAN}[3]{#2}
\let\VANthebibliography\thebibliography
\def\thebibliography{\DeclareRobustCommand{\VAN}[3]{##3}\VANthebibliography}
\usepackage{graphicx}
\usepackage{xcolor}
\usepackage{enumitem} 
\usepackage{enumitem}
\newcommand{\beq}{\begin{equation}}
\newcommand{\eeq}{\end{equation}}
\def\lap{\lower.5ex\hbox{$\; \buildrel < \over \sim \;$}}
\def\gap{\lower.5ex\hbox{$\; \buildrel > \over \sim \;$}}

\usepackage[utf8]{inputenc}

\title{Anomalies in Physical Cosmology}
\author[P. J. E. Peebles]{P. J. E. Peebles,$^{1}$\thanks{E-mail: pjep@Princeton.edu}\\
$^{1}$Joseph Henry Laboratories, Princeton University, Princeton, NJ 08544, USA\\
pjep@princeton.edu
}

\begin{document}

\maketitle 
\begin{abstract} 
The $\Lambda$CDM cosmology passes demanding tests that establish it as a good approximation to reality. The theory is incomplete, of course, and open issues are being examined in active research programs. I offer a review of less widely discussed anomalies that might also point to hints to a still better cosmological theory if more closely examined.
\end{abstract}

\section{\bf Introduction}\label{intro}

\noindent{\it The world is so full of a number of things,}

\noindent{\it I'm sure we should all be as happy as kings.}

{\qquad\qquad\qquad\qquad Robert Louis Stevenson}\medskip

\noindent In the empiricist philosophy of this essay\footnote{This is a much revised version of the unpublished sentiments expressed in the draft paper, Peebles (2021).} experiments and observations have two key functions: the puzzles they reveal inspire theories, and the theories are judged by the degree of empirical success of their predictions. The relative importance of the two functions can be quite different in different cases. Phenomena  played a small though certainly significant role in forming Einstein's intuition about a philosophically satisfactory theory of gravity, the general theory of relativity. It has been said that general relativity was ``there, to be recognized,'' essentially by pure thought. But in the empiricist philosophy Einstein's great accomplishment is to be particularly celebrated because the theory Einstein found passes many tests that make the empirical case that general relativity is a remarkably useful approximation to the reality whose nature we seek to discover. Phenomenology played a much greater role when Maxwell gave up seeking a mechanical model for the ether and took the intuitive step of completing the field equations that pass such broad empirical tests. Quantum physics, and the standard model for particle physics, grew in an intermediate manner, a combination of brilliant ideas and helpful phenomenology. The same is true of our present physical cosmology, which grew out of suggestive phenomenology and intuitive leaps, some of which failed while others prove to be remarkably successful. The result, the $\Lambda$CDM theory, agrees with a broad range of well-checked predictions. But the thought that there is an even better classical cosmology to be found is encouraged by intuition, as in the feeling that the stark simplicity of the dark sector of the theory surely is inadequate, and from curious issues in the phenomenology, which is the topic of this paper. 

To reduce the chance of misunderstanding I emphasize that the empirical case that the  $\Lambda$CDM theory is a good approximation to reality remains compelling. But I argue in this paper that we have empirical evidence that there is a still better theory to be found. 

The goal of an improved cosmology, and the far greater great goal of a full reconciliation of the quantum and relativity principles, might be approached by strokes of insight, following Einstein's example, or by incremental advances, following Maxwell: searches for improvements inspired by anomalies. By this I mean empirical evidence that seems likely to disagree with what is expected from standard and accepted theories and ideas. An anomaly might prove to be only apparent, resolved by improved empirical evidence or a better appreciation of the predictions of the theory we already have. Past experience suggests that others will prove to be real, and will be a valuable stimulus to the exploration of new or previously neglected ideas about physical theory. This includes cosmology, and maybe even unification of the basic principles of physics. But a final theory, if there is such a thing, cannot be empirically established, only checked to the extent that world economies can afford. 

Another consideration follows from the phenomenon of multiples in scientific discovery: apparently independent appreciations of the same idea (as reviewed by  Merton 1961). Examples are common; examples in the development of the present standard cosmology  are reviewed in Peebles (2020a; 2022a). Some are results of advances in technology or theory that suggest ideas that are in the literature, and could be appreciated by more than one person. Other examples are results of communication by many means, including  nonverbal hints to directions of thinking, which can cause growing community recognition of ideas that are ``there, to be recognized,'' in previously neglected phenomena. This invites the hopeful thought that there are still other phenomena that might be recognized to be useful hints to the improvement of physics, if examined more closely. The thought is particularly relevant for cosmology because the universe has room for an immense number things. I argue that some seem to me to call for closer attention. 

Research programs tend to pursue systematic explorations of scientific issues that the community agrees are important. An example in cosmology is the issue of cores and cusps in the distributions of dark and baryonic matter in the centers of galaxies. The issue is important, a guide to the establishment of constraints on the nature of dark matter and the theory of galaxy formation. Research in progress is examining the theory and observations. This is good science: close attention to a specific question might reveal something interesting. But it is good science also to cast about for more obscure issues that seem curious but have not yet been examined as carefully as might be useful. My purpose in this essay is to offer thoughts about anomalies in cosmology that are not so widely discussed, and might aid advances in this subject if given more attention. 

Section~\ref{sec:anomalies} is a commentary on more familiar anomalies in physical science that seem relevant to cosmology. The subjects of the next three sections are anomalies, real or apparent, that are less well advertised for the most part. Section~\ref{sec:distributions}  reviews open issues in the large-scale distributions of radio galaxies, AGNs and clusters of galaxies. The curious properties of the Local Void are considered in Section~\ref{sec:localvoid}, and apparent anomalies in the properties of galaxies are reviewed in Section~\ref{sec:galaxies}. Section~\ref{SummaryRemarks} presents an overview of what I consider to be the main points of this essay. 

We must consider first the reliability of the basis by which phenomena are judged to be anomalous within accepted thinking. The empirical case that the relativistic hot big bang $\Lambda$CDM cosmology is a useful approximation to reality is outlined in Section~\ref{sec:EmpiricalBasis}. This theory is not exact, but it does not seem likely to be far off. Thus my assessments of issues and anomalies in cosmology take it that the tests have persuasively established that the $\Lambda$CDM theory is a good approximation to reality that likely requires refinement.

The anthropic principle is another consideration by which theories are motivated and their anomalies evaluated and maybe resolved. The line of thinking has been particularly influential in research in physical cosmology, and so deserves mention here. What is more, we must expect that social constructions, aided by the anthropic principle, will be forced on the scientific community eventually, if the research community survives that long, as theory outstrips the ability to test predictions. But I argue in Section~\ref{sec:Anthropic} that the anthropic approach is of doubtful use for the present purpose because we have immense room between the extremes of the unapproachably large and unapproachably small scales for continued empirical exploration.

My citations to the literature are limited to a few introductory papers and recent review articles, which I hope aids readability while indicating the present state of thinking about the science and the evidence. I do not comment on issues in cosmology where I have no thoughts to add to what already is in the literature.\footnote{\label{fn:S0s}Choices of interesting issues to explore in what we aim to be the advance of objective science must be subjective. The standard cosmology predicts a small but important primeval abundance of the lithium isotope $^7$Li. The prediction is clear, but I am not comfortable assessing the astrophysics of production and depletion of lithium. Planar alignments of satellites of galaxies might be significant, but I worry about our powerful ability to see patterns even in noise. Thinking about the future of the universe, maybe a big crunch or big rip, offers great adventures of the mind, but it is only empirically interesting if there is something to observe, maybe remnants of the last phase of a cyclic universe. I am particularly uneasy about my lack of attention to the S0 galaxies that are common in clusters, and present but not common among nearby galaxies. An S0 looks somewhat like a spiral galaxy whose arms have been erased leaving a disk of stars usually with a relatively large stellar halo. S0s may have readily interpretable things to teach us about cosmic structure formation, but I have not grasped them.} The declarative sentences in this essay are my opinions. For other reviews of independent selections of issues and anomalies in cosmology I refer to Abdalla, Abell{\'a}n, Aboubrahim, et al. (2022) and Perivolaropoulos and Skara (2022). A different philosophy informs the assessment of the empirical situation in cosmology by Subir Sarkar (2022). 

\subsection{Empirical Basis for the Standard Cosmology}\label{sec:EmpiricalBasis}

To help clarify discussions of tests and anomalies we need a definition of the standard $\Lambda$CDM cosmology. The version used in this paper is discussed in Section~\ref{sec:definition}. Section~\ref{sec:tests} outlines the tests that establish this cosmology as a useful approximation to reality.

\subsubsection{Definition}\label{sec:definition}

Einstein's cosmological principle, or assumption, is that the universe is close homogeneous and isotropic in the large-scale average.  To be more explicit about the role of this assumption in the standard $\Lambda$CDM theory used in this paper I offer the following definition. The theory applies the standard physics of matter, radiation, and Einstein's general theory of relativity with its cosmological constant to a cosmologically flat universe that is a spatially stationary, isotropic,  random process with a close to scale-invariant power law power spectrum of  Gaussian and adiabatic departures from homogeneity.  This trimmed-down theory has eight free parameters (the density parameters in ordinary matter, dark matter, the CMB, and neutrinos with negligible rest masses; with  Hubble's constant, the primeval Gaussian process amplitude and power law index, and the optical depth for scattering of the CMB by intergalactic plasma). 

The mathematician and statistician Jerzy Neyman (1962) pointed out that the rational statement of Einstein's cosmological principle is that the universe is assumed to be a realization of a ``stationary stochastic process,'' or as it is put here a stationary random process. The process is what a theory with given parameters is supposed to predict. A measurement to test the theory and constrain the parameters uses a sample, which is considered a realization of the process. A fair sample is a realization that has statistical properties that are usefully close to the predictions of the process. Analyses of what is arguably a fair sample test whether the theoretical process is an adequate approximation to reality. 

Neyman remarked that the empirical evaluation of a deterministic classical theory is necessarily indeterministic, because no finite sample can determine parameters to arbitrarily tight precision. To my knowledge Neyman was the first to apply this thinking to our strictly limited sample of the universe, our sample of the stationary random process of the cosmological principle. We can add another reason for the  indeterministic nature of physical science: our theories are incomplete. A notable example is the inconsistency of the quantum and relativity principles (Sec.~\ref{sec:QM&GR}).

To make progress we must add to the definition of the $\Lambda$CDM theory the less specific provision that the realization of the random process that is our observable universe is close enough to a fair sample to allow reliable computations of testable predictions. Aspects of this situation are discussed in Section~\ref{sec:theacausaluiverse}. A related provision is that the primeval power law power spectrum of the primeval mass distribution must be truncated to avoid a divergence of spacetime curvature fluctuations. The measured scalar spectral index, $n_s\simeq 0.96$, is associated with primeval spacetime curvature fluctuations that scale with size $r$ as $\delta\phi\propto r^{(1-n_s)/2}\sim r^{0.02}$. The power spectrum has to be truncated at some large scale, but it will be assumed that the wavenumber at the truncation doesn't much matter.  Variants of this standard picture, such as a Tilted Universe (Turner 1991), in which we might not have a fair sample, are not explicitly considered here. 

Other authors use different definitions of the cosmological principle.  Secrest, von Hausegger, Rameez, et al. (2022) take the principle to be ``that the universe on large scales must appear to be the same to all observers, independent of their location.'' I am indebted to Subir Sarkar for pointing out that Milne (1933) presented essentially the same definition. This is of historical interest because Milne introduced the term, Einstein's cosmological principle, and pointed out that Hubble's law follows from this principle without application of any deeper theory. But consider that fluctuations in the distributions of extragalactic objects on scales we can observe within the present Hubble length were at earlier epochs not observable. Why should there not be density fluctuations beyond the present Hubble length that are observable if at all only through their indirect effects? Neyman's philosophy accommodates this. Still more aspects of the situation are discussed in Section~\ref{sec:theacausaluiverse}.

\subsubsection{Cosmological Tests}~\label{sec:tests}

Surveys of the tests that make the case that the $\Lambda$CDM cosmology is a useful approximation to reality, and that an even better theory will look much like $\Lambda$CDM, are widely available in the literature. My contribution is in Peebles (2020a). I offer here a reminder of key points in my positive assessment of the situation. 

The measured statistical patterns in the space distribution of the galaxies and in the angular distribution of the thermal cosmic microwave background radiation, the CMB, agree with what is expected from the remnants of acoustic oscillations of the plasma and radiation that acted as a fluid up to recombination at redshift $z\sim 1000$.  The precision of the CMB anisotropy measurements and the tight consistency with the $\Lambda$CDM predictions is deeply impressive. But even more impressive is the consistency with the theory and observation of the pattern in the space distribution of the galaxies that in theory also is a remnant of the acoustic oscillations. This cosmological test of consistency is based on two quite different phenomena, the space distribution of the galaxies observed at redshifts less than unity and the pattern in the angular distribution of the CMB that in theory was formed at redshift $\sim 1000$. The patterns in the two distributions are measured by different methods of observation and data reduction, yet they agree with the same $\Lambda$CDM universe. This consistency is a demonstration that we have a good approximation to reality that is as close to convincing as we can get in natural science.

Other tests of consistency, in examples based on what happened over a broad range of redshifts, add weight to the case for the $\Lambda$CDM theory. The values of the cosmic mean mass densities in baryons and dark matter, the helium mass fraction, and the cosmological constant that are needed to fit the CMB anisotropy measurements that probe the state of the universe at $z\sim 1000$ agree with the baryon density that fits the formation of the isotopes of hydrogen and helium at $z\sim 10^9$, the abundances of helium in the Sun, interstellar plasma, and planetary nebulae,  the dynamical mass density derived from relative motions of the galaxies at redshifts less than unity, the mass density and cosmological constant derived from the supernova redshift-magnitude relation at $z\sim 1$, the stellar evolution ages of the oldest known stars, and the angular size distance as a function of redshift at $z\lap 1$. 

There are discrepancies, real or apparent. The values of Hubble's constant, $H_{\rm o}$, that are needed to fit the CMB anisotropy measurements differ from the relation between distances and recession speeds of relatively nearby galaxies by about 10\%. This well-discussed Hubble tension, if real, is a 10\% error arising from tracing the expansion of the universe to the present by a factor of a thousand from the epoch of formation of the patterns in the distributions of the baryons and CMB. I count this as an impressive success to be added to the rest of the evidence that the $\Lambda$CDM theory is a useful approximation to reality, though of course not exact. If the anomaly in the two measures of $H_{\rm o}$ is real then surely other anomalies are to be found. Another tension is the evidence that  the normalization of the mass fluctuation power spectrum required to fit the CMB anisotropy measurements differs from the normalization required to fit measurements at low redshifts: gravitational lensing, the galaxy two-point correlation function in redshift space, and counts of clusters of galaxies. The normalizations, measures of $\delta M/M$, again differ at high and low redshifts by 10\% (in the compilation by Perivolaropoulos and Skara 2022). 
   
Einstein's general theory of relativity passes tight tests on the scale of the Solar System down to the laboratory, on scales  $\lap 10^{13}$~cm. Cosmology applies this theory at distances on the order of the Hubble length, $10^{28}$~cm. As a general policy, would you trust an extrapolation of a theory by fifteen orders of magnitude in length scale? But we have a prior example, the enormous range of scales of the successful applications of quantum physics, from superconductors to detection of the higgs boson. The success of the cosmological tests we have so far gives considerable weight to this great extrapolation of general relativity to its application in the $\Lambda$CDM theory. 

The key point from these considerations is that we have a broadly consistent story from a considerable variety of ways to observe the nature of the universe by different methods of observation and analysis, by groups that are operating independently and, it is reasonable to expect, interested in finding anomalies that may prove to be interesting. It would be ridiculous to suppose that the network of tests is wrong or misleading but has converged to apparent consistency by some combination of accidental and unintended errors, let alone conspiracies. We have instead excellent reason to expect that a better theory to be discovered will look a lot like $\Lambda$CDM, because the $\Lambda$CDM universe has been shown to look a lot like what is observed. This is the basis for the conclusion that we have a useful approximation to reality, and the hope that empirical anomalies will offer hints to improvements.

The search for improvements includes exploration of the consequences of modified gravity physics. It would be a curious coincidence if the theory required modification just on reaching the scales of cosmology, however. A better immediate prospect for improvement is the physics of the dark sector of $\Lambda$ and dark matter, which seems artificially simple. 

We cannot prove as a theorem that no other physical theory could predict the reasonable degree of consistency of theory and observation of these many different ways to probe the universe; we do not do theorems in natural science. But we can conclude from these tests that we have a compelling case that the $\Lambda$CDM theory is a useful approximation to reality, and we can hope to see improvements of the theory as the observations improve. 

\subsection{The Anthropic Principle}\label{sec:Anthropic}

If society continues to be willing and able to support curiosity-driven research in the natural sciences there will come a time when physicists have found a final theory of everything that is internally consistent and agrees with all available tests, but the assessment by tests of predictions will be impossible. That would be impossible in principle, if something like a multiverse is involved, or impossible in practice, because the world economy cannot afford tests of the predictions of this final theory. It will be reasonable and sensible  to consider this theory to be a persuasive nonempirical establishment of reality (Dawid 2013). But it will be sensible also not to be quite sure of what must be a social construction (Peebles 2022). 

We have a precursor to this dilemma, the anthropic principle. It offers one way to deal with an anomaly: postulate an ensemble of all possible universes and observe that we could flourish only in one suited to our needs expressed so as to account for the anomaly. Reactions to this line of thought differ. Some dismiss it as a ``just so'' story. Steven Weinberg (1989) pointed out that it is one way to account for the quantum vacuum energy density, which looks likely to be quite unacceptable large. This section is meant to explain my feeling that the anthropic principle is not an appropriate guide to considerations of anomalies in physical cosmology. 

Robert Henry Dicke (1961) introduced a weak form of this argument. Dicke pointed out that  the universe has to have been expanding for at least a few gigayears. The time is needed to allow for the evolution of several generations of stars that produced the heavy elements we need, then the formation and cooling of the solar system, and then the evolution of the species up to observers who take an interest in the expanding universe. This is a consistency condition. Better put, it is the assumption that Nature abhors logical inconsistencies. 

An argument based on a more adventurous form of the anthropic principle starts from the evidence that there are enormous numbers of planets around stars in our galaxy. This allows room for many planets capable of hosting beings similar to us. The frequency distribution in cosmic times when these beings flourish on different planets might be expected to peak at about $10^{10}$~yr, because this allows time for natural evolution while avoiding the serious slowing of star formation at much greater cosmic times. This time is about what is observed on our planet; we flourish about when might be expected. An empirical test from a modest sampling of what is on nearby planetary systems might be possible, eventually. 

 Weinberg (1989) discussed a stronger form that postulates a statistical ensemble of universes, a multiverse. If, for example, universes in the ensemble that have shorter expansion times are more numerous, then the odds are that we live in one of the universes with the minimum expansion time consistent with what is required to allow our existence, which seems about right. Weinberg applied this thinking to the curiously small value of the cosmological constant compared to what is expected from quantum physics. Weinberg postulated that the laws of physics in each universe in the ensemble would be different. We could only flourish in a universe with physics similar to ours on the level we require, but that degree of similarity could allow a broad spread of values of the quantum vacuum energy density, $\Lambda$, provided that that depends on deeper physics that does not affect our well-being. There would be universes in the multiverse that satisfy this condition and the value of $\Lambda$ is not so negative that the universe stops expanding and collapses too soon for the span of time we required, and not so positive and large that the rapid expansion driven by $\Lambda$ would have prevented the gravitational assembly of galaxies. If galaxies in the ensemble that have larger absolute values of $\Lambda$ are more common, as might be expected from the large value expected of the quantum vacuum, then we would expect to find ourselves in a universe with a value of $\Lambda$ that is about as large as is consistent with our existence. This is about what is observed. 

Martin Rees (2020) rightly celebrates the concept of the multiverse as the next layer in the sequence of revolutions in our understanding of the nature of the world around us. Ideas about this have passed through many layers: the Ptolemy universe with the earth centered in the crystal spheres that hold the astronomical objects; the Copernican universe with the sun at the center; the Kapteyn universe centered on the Milky Way galaxy; Hubble's realm of the nebulae with no center; and the multiverse of which our universe is but a speck. The layers of discovery go down in scale too. Henri Poincar\'e (1902) remarked that the Mariotte/Boyle law is wonderfully simple and accurate for many gases, but these gases examined in sufficiently fine detail break up into the complex motions of enormous numbers of particles. Poincar\'e asked whether gravity examined in sufficiently fine detail might also depart from the simplicity of Newton's law into complex behavior. Poincar\'e suggested we consider that ``then again [there may be] the simple under the complex, and so on, without our being able to foresee what will be the last term.'' Maybe underlying the particle physicists' concept of a theory of everything are yet more layers of Poincar\'e's successive approximations. And why should the layers of structure on large scales not continue to multiverses and beyond? 

 Weinberg (1989) cautioned that the anthropic upper bound on the absolute value of $\Lambda$ is well above the bound that could be set by astronomical observations we had then. There are other examples of what seem to be excessive satisfaction of the anthropic condition. We need at least one gravitational potential well similar to that of a galaxy to contain and recycle the debris from a few generations of stars to have produced the chemical elements we require. But did we need the observed enormous number of galaxies? Would the already large number of planetary systems among the $\sim 10^{11}$ stars in the Milky Way have been adequate? If the Milky Way would serve, and given that it is present, does physics require all those other galaxies? In implementations of cosmological inflation the numbers of galaxies, and their sizes and densities, depend on the amplitude of the primeval fluctuations in spacetime curvature associated with the primeval departures from an exactly homogeneous mass distribution. A universe identical to ours except that the fluctuations in spacetime curvature are an order of magnitude smaller would develop far fewer galaxies that are less dense, but would that be a problem for our existence? We have so many galaxies to spare. If in the multiverse the universes with smaller primeval curvature fluctuations were more common then application of the anthropic consideration would lead us to expect far fewer galaxies than observed; we don't need so many. If universes with larger primeval curvature fluctuations were more common we would expect to find ourselves in an accidental island of tranquility among the chaos of violent mergers and relativistic collapses. Again, neither situation is observed.

Weinberg's argument is good science; it explores a possible aspect of reality that accounts for the great difference between the value of the $\Lambda$ of cosmology and the value expected from quantum physics. But there are troubling aspects of this approach. The excess of baryons over antibaryons in the Local Group could be attributed to the anthropic principle: we live in a universe drawn from the multiverse that has the excess baryon density acceptable for our existence. But theories of particle physics and physical conditions in the early universe might predict the excess. Given the choice of seeking this better physics or resorting to the anthropic principle, I expect most would choose the former. It can be awkward to base arguments on the Panglossian principle that we inhabit the best of all possible universes suited to our existence.

\section{Anomalies in Physical Science}\label{sec:anomalies}

Some anomalies that have long resisted interpretation are so familiar that they tend to pass without mention. An example is Wigner's (1960) ``Unreasonable Effectiveness of Mathematics in the Natural Sciences.''

\subsection{Physics is Successful} \label{PhysicsisSuccessful}

Eugene Paul Wigner (1960) wrote about the ``two miracles of the existence of laws of nature and of the human mind's capacity to divine them.'' Natural scientists are conditioned to accept these two miracles, or phenomena, as self-evident; they are essential for the discoveries of well-tested science that makes possible the vast range of technology we all experience.  But we should pause on occasion to consider that Wigner's  two miracles are assumptions. Experience supports them, but of course never offers a proof.

The starting miracle to add to these two is that the world around us exists, and has existed in some form or another for a very long time, with the properties one would expect of physical reality. In natural science we must take this as given. We assume then Wigner's two miracles and notice that they satisfy two consistency conditions. First, if macroscopic physics were not reproducible and lawful it would be difficult to imagine the natural evolution of the species: what is the use of adapting to physical properties of matter that can change on time scales less than cosmic? Second, if it were not possible to discover useful approximations to the laws of nature then I suppose we would not be marveling about it. These thoughts could be taken to suggest that the assumption of lawful behavior, which is such a good approximation, might be found to fail as we probe ever deeply into the nature of the world, at levels where erratic behavior need not have had a deleterious effect on living matter. It calls to mind quantum physics. 

\subsection{The Relativistic and Quantum Principles are Inconsistent}\label{sec:QM&GR}

In the empiricist philosophy the use of quantum physics to account for the spectra of galaxies at redshift $z=10$ and the formation of the light isotopes at $z\sim1000$, applied in the strongly curved spacetime of relativity, need not be problematic even though the quantum and relativity principles are not consistent.  Maybe this is a signal of an intrinsic inconsistency; maybe physics is not exactly lawful. The improbably large estimate of the quantum vacuum mass density discussed in Section~\ref{sec:Lambda} is a real problem. More abstract, but a worry, is whether the quantum physics of observables operating on state vectors is a useful approximation when extrapolated to describe the whole universe. Would a quantum universe in a pure state really decohere into the classical world of general relativity? Would a viable theory of our universe taken to be a mixed state be any different from our present theory? A reasoned assessment of such issues is beyond my ability and the scope of this essay. 


\subsection{The Symmetry of Matter and Antimatter is Broken}\label{sec:baryonnumber}

A familiar anomaly that is essential to our existence is the pronounced local excess of matter over antimatter. The sensitive tests for anti-helium by the Alpha Magnetic Spectrometer indicate the Milky Way contains little antimatter (Poulin, Salati, Cholis, Kamionkowski, and Silk 2019; Aguilar, Ali Cavasonza, Ambrosi, et al., 2021). We can add the evidence from the absence of detection of gamma ray annihilation radiation from dwarf  galaxies that have plunged into the Milky Way. And since satellites of the Milky Way and our neighbor M~31 surely have intermingled before falling into one or the other of the large galaxies without producing detectable gamma rays it seems clear that the Local Group is made of baryons, with a tiny fraction of antibaryons. 

Should we revisit the question of whether some galaxies are made of anti-baryons? A sharp division of regions of matter and antimatter could invite an unacceptably large surface mass density in domain walls, but an imaginative physicist might find a way around that. The literature on possible extensions of the standard model for particle physics and the physical conditions that would account for baryogenesis continues to grow. I am not competent to review the state of the art.

\subsection{Should Local Physics be Evolving?}\label{sec:EvolutionPhysics} 

We have been given leave by string theory to imagine that the dimensionless parameters of physics are evolving, because in the varieties of string theory it is difficult to see what fixes their values. Thus Uzan (2003, Sec.~VI.B) concludes that ``as yet no complete and satisfactory mechanism for the stabilization of the extra dimension and dilaton is known," and with it stabilization of the dimensionless parameters of fundamental physics. This complements an older thought based on a measure of the strength of the gravitational interaction, 
\beq
{\cal G} = {Gm^2\over \hbar c}\sim 10^{-38}, \label{gravitystrength}
\eeq
where $m$ is the mass of a nucleon. This number is remarkably small, one might say anomalously so. An anthropic explanation is that if ${\cal G}$ were much larger it would make stellar evolution times too short to allow for evolution of the species by natural selection, even on a suitably small planet near a suitably low mass long-lived star. 

Dirac (1937) argued that it is difficult to imagine how the tiny value of ${\cal G}$ could follow from a fundamental theory that might be expected to produce numbers such as $\pi$ and $e$, and integers of modest size and their fractional powers. Dirac suggested that a hint might be drawn from the comparison of ${\cal G}$ to the ratio of the atomic time $e^2/mc^3$ to the Hubble expansion time, then estimated to be $t \sim 2\times 10^9$~yr, giving the dimensionless number
\beq
{\cal T} = {e^2\over t~mc^3} \sim 10^{-38}. \label{atomictime}
\eeq
It is curious that the two exceedingly small numbers, ${\cal G}$ and ${\cal T}$, are similar. The value of ${\cal T}$ is decreasing, assuming the universe is evolving and local physics is not. Maybe ${\cal G}$ is small, and comparable to ${\cal T}$, because ${\cal G}$ is evolving to its natural value, zero, along with ${\cal T}$. 

The measure of the electromagnetic interaction,
\beq
\alpha = {e^2\over\hbar c}\sim {1\over 137}, \label{eq:alpha}
\eeq
is not such a small number. But if ${\cal G}$ is not constant then $\alpha$ surely need not be fixed either.

Einstein's cosmological constant, $\Lambda$, represents the effective vacuum energy density, 
\beq
\rho_\Lambda = {3 H_{\rm o}^2(1 - \Omega_m)\over 8 \pi G}
\sim 10^{-29}$~g~cm$^{-3}.
\eeq
The ratio of this quantity to the Planck mass density, $\rho_{\rm Planck}=c^5/\hbar G^2$, is
\beq
{\cal L} = {\rho_{\Lambda}\over\rho_{\rm Planck}} \sim 10^{-123}.
\eeq
The empirical evidence is that $\cal L$ is not zero (as reviewed in Secs.~\ref{sec:tests} and \ref{sec:Lambda}), but rather this dauntingly small number. Maybe it calls for application of the anthropic principle, as Weinberg (1989) argued. Or, following Dirac's thinking, maybe this number is so small because local physics has been evolving for a long time, and with it the value of $\cal L$. 

Dicke was fascinated by the thought that the laws of physics might change as the universe evolves. My impression is that this was at least in part because checking on the idea led to fascinating explorations of the many possible sources of empirical evidence for or against evolution, drawn from the laboratory, geology, astrophysics, and on to cosmology, always with close attention to the empirical content (e.g. Dicke 1964). My PhD dissertation under his direction was on the theory and empirical constraints on the evolution of the strength $\alpha$ of the electromagnetic interaction. Dicke played an important role in setting up the Lunar Laser Ranging experiment that has yielded  tight tests of physics, including the evolution of ${\cal G}$. The experiment has established that at the present epoch the strength of the gravitational interaction is not evolving faster than about one percent of the rate of expansion of the universe (Williams, Turyshev, and Boggs 2004). Dicke's program led Bharat Ratra and me to explore the thought that the value of the cosmological constant $\Lambda$ is so small compared to what is expected from quantum physics because $\Lambda$ is not a constant: it has been slowly evolving to its only reasonable value, zero. Ratra and Peebles (1988) present this thought and a model in which it happens. 

 The search for evidence of evolution of the effective value of the vacuum energy density $\Lambda$ under its new name, dark energy, is now widely discussed and well supported; an example is the Dark Energy Survey, DES (DES Collaboration 2022). It is equally good science to test whether the strength of the electromagnetic interaction, $\alpha = e^2/\hbar c$, might evolve. Advances in technology to test this are discussed in recent papers by  Murphy, Molaro, Leite, et al.(2022) and Webb, Lee, and Milakovi{\'c} (2022). So why isn't there a well-supported Fine-Structure Survey, FSS? 

The challenge posed by Dirac (1937) and Dicke (1964) remains: discover what accounts for the measure of the strength ${\cal G}$ of the gravitational interaction in equation~(\ref{gravitystrength}) that is so far from what might be expected to be predicted by a theory of everything. A challenge of the same sort is to account for the great difference between the values of the effective vacuum energy density $\Lambda$ of cosmology and the Planck mass density suggested by quantum physics (Sec.\ref{sec:Lambda}). Maybe these  pronounced anomalies will be explained by subtle aspects of a deeper theory that predicts the tiny values of ${\cal G}$ and $\cal L$. Maybe we must resort to the anthropic philosophy. Or maybe both ${\cal G}$ and $\cal L$ have been decreasing, ${\cal G}$ following a course of evolution that happens to have escaped the tight constraint from the Lunar Laser Ranging experiment. 

\subsection{The Standard Cosmology is Singular}

The $\Lambda$CDM theory defined in Section~\ref{sec:definition} predicts that the expansion of the universe traces back to a singular state of arbitrarily large density. It is usually supposed that Nature abhors singularities, and that this one necessarily points to incompleteness to be remedied by a better theory. The most widely discussed possible extension of $\Lambda$CDM is cosmological inflation (Guth 1981). It removes the singularity, but Borde, Guth, and Vilenkin (2003) argue that even eternal inflation cannot have been eternal back to the arbitrarily remote past. This is not an argument against inflation, eternal or otherwise, only that the term, eternal, cannot be the whole picture. We make progress by successive approximations.

When the concept of cosmological inflation first gained general attention an  implication was taken to be that space sections are flat with close to Gaussian adiabatic departures from homogeneity and a power law power spectrum slightly tilted from scale-invariance. This was before each of these conditions were  observationally established, and it no longer matters whether inflation really predicts all this. In the empiricist philosophy of this essay this history makes inflation particularly interesting, though of course the empirical case is not yet persuasive because tests of predictions are still scant.

\subsection{The  Standard Cosmology is Acausal}\label{sec:theacausaluiverse}

Hubble's (1936) ``Realm of the Nebulae,'' the observed galaxies, does not have a noticeable edge. This is consistent with the $\Lambda$CDM theory defined in Section~\ref{sec:definition}, but it is an anomaly because in this theory distant galaxies observed in well-separated parts of the sky have not been in causal contact no matter how far back in time the expansion is traced, to the singularity. So how did the galaxies ``know'' how to resemble each other? 

The resolution offered by the cosmological inflation picture is that there was a time in the early universe when a near exponential rate of expansion produced a far larger horizon, resulting in causal connection across all we can see. The successful empirical tests of cosmology outlined in Section~(\ref{sec:tests}), which depend on this acausality, are in turn evidence of the effect of inflation in some form. At the time of writing, however, we have only a modest empirical basis for assessments of specific theories of this aspect of what happened in the remote past. 

In the standard theory the gravitational growth of primeval departures from an exactly homogeneous mass distribution, which is discussed in Section~\ref{sec:localgravity}, is acausal, as follows. Let the universe be cosmologically flat, ignore the mass in radiation and the pressure of matter, and assign time-orthogonal coordinates that eliminate the decaying mode of the departure from a homogeneous mass distribution. Then in linear perturbation theory the motion of the matter relative to the general expansion of the universe is
\beq
v^\alpha(\vec x,t) = {Ha(t)f(\Omega)\over 4\pi}
{\partial\over\partial x^\alpha} \int d^3x'{\delta(\vec x',t)\over |\vec x' - \vec x|}. 
\label{eq:peculiaracceleration}
\eeq
Here $H$ is Hubble's constant, $a(t)$ is the expansion parameter, $f\simeq \Omega^{0.6}$ where $\Omega$ is the density parameter, and $\delta(\vec x,t)$ is the fractional departure from a homogeneous mass distribution. We see that in the gravity physics of the standard cosmology a mass concentration $\delta(\vec x',t)$ that is in principle too far away to be observed can produce a flow $v^\alpha(\vec x,t)$ that in principle can be observed. Grishchuk and Zel'dovich (1978) may have been the first to recognize this. 

The situation illustrated in equation~(\ref{eq:peculiaracceleration}) is acausal, but we have learned to live with it by placing it in the initial condition that the universe is a stationary random process. This means that in the early universe the cosmic structure we observe would have been waves in the mass distribution with tiny amplitude and spread across far broader lengths than the particle horizon. Put another way, in standard $\Lambda$CDM the statistical homogeneity of the universe and its primeval power spectrum of departures from homogeneity have acausal origins. 

Anther aspect of the acausality of our present cosmology arises in the theory and observation of the angular distribution of the thermal cosmic background radiation temperature $T(\theta, \phi)$ as a function of angular position across the sky. It is conveniently expressed as the spherical harmonic expansion 
\beq
T(\theta, \phi) = \sum a_\ell^mY_\ell^m(\theta, \phi). \label{eq:harmonicexp}
\eeq
In the standard theory the real and imaginary parts of the expansion coefficient $a_\ell^m$, if measured in many different realizations of the universe, would have gaussian distributions around zero mean. On small scales, large degree $\ell$, there are in effect many observations of realizations of the random process across the sky, so the measurements of the absolute values $|a_\ell^m|^2$, averaged over $m$, have little scatter in the prediction and  measurement. This allows the impressively close checks of the $\Lambda$CDM theory over a large range of values of $\ell$. The situation is different at small $\ell$, large angular scale, because the $|a_\ell^m|^2$ are averaged over only a few $m$. The result is an uncertain measurement to be compared to an uncertain prediction that depends on what is in parts of the universe we cannot observe in principle.

\subsection{Why Dark Matter?}\label{DM} 

The starting idea for nonbaryonic dark matter was that it might be a new family of neutrinos with the standard V-A coupling to their leptons and neutrino rest mass $\sim 3$~Gev. The mass was chosen so the remnant abundance from the thermal production and annihilation of these hypothetical neutrinos in the hot early universe would be interesting for cosmology. It was introduced, independently as far as I know, in five papers: Hut (1977); Lee and Weinberg (1977); Sato and Kobayashi (1977);  Dicus, Kolb, and Teplitz (1977);  and Vysotskij, Dolgov, and Zel'dovich (1977). This candidate for dark matter has become known as weakly interacting massive particles, or WIMPs. Discussions of possible detection of WIMPs began with Steigman, Sarazin, Quintana, and Faulkner (1978), who pointed out that interactions of a sea of massive neutrinos with themselves and the baryons might allow accumulation of these neutrinos in stars and planets, with observable effects, as in what became known as dark stars. Cold dark matter was more formally added to the relativistic cosmological model by Peebles (1982), for the purpose of showing how the smooth distribution of the CMB can be reconciled with the clumpy distribution of the galaxies in the picture of gravitational formation of cosmic structure. 

The first laboratory attempts to detect interactions of WIMPs with ordinary matter began in the 1980s. Schumann (2019) reviews the considerable advances in sensitivity since then in a considerable variety of experimental designs, mass scales, and effective isolation from cosmic rays and other local noise. There are other candidates for dark matter, with active research aimed at their detection. Examples are the lightest supersymmetric partner, axions, fuzzy dark matter, supersymmetric dark matter, and black holes. The searches for signs of these objects and more, in the laboratory and astronomical observations, have been energetically pursued for four decades, and at the time of writing they have not yielded a generally accepted detection of any form of dark matter (apart from the small contributions by the neutrinos in the three known lepton families). Several aspects of this situation are to be considered. 

First, the absence of detection despite many years of great effort does not falsify the assumption of nonbaryonic dark matter in the $\Lambda$CDM theory. This dark matter is an essential postulate, but the theory places no requirement on its place in an extended standard model for particle physics. More broadly put, we must bear in mind that we have no guarantee that we can discover how to fit well-established phenomena into satisfactory theories. The success so far in turning anomalies into physical theories that pass challenging tests has been so productive that the failure to detect dark matter despite considerable effort might be considered anomalous on historical grounds. But to repeat: the failure of detection would conflict with experience in physics, but it would not conflict with the $\Lambda$CDM theory. 

Second, we now have in physics two types of matter, baryonic and dark. Properties of the former are known in great detail; properties of the latter are only roughly constrained. Ideas about how baryonic matter formed look promising but have not yet converged on a generally accepted theory (Sec.\ref{sec:baryonnumber}). Ideas about how dark matter formed must be fragmented because they depend on how dark matter fits the rest of particle physics, which is not known. Since dark matter does not seem to be necessary for our existence the Panglossian philosophy of the anthropic principle discussed in Section~\ref{sec:Anthropic} would have it that the creation of dark matter must necessarily have accompanied the creation of the baryonic matter we certainly need. It could account for the otherwise curious coincidence of the comparable amounts of baryonic and dark matter, as in asymmetric dark matter (e.g., Zurek 2014).

Third, the elegance and predictive power of the rich physics of known matter inspires the thought, or hope, that the dark sector surely is more interesting than a gas of freely moving particles with initially small velocity dispersion along with a constant mass density that has a distinctly odd value. It makes sense to explore the idea that the physics of the dark sector resembles elements of our established particle physics. Influential examples include the Sommerfeld/Coulomb enhancement of scattering by a Yukawa potential, in the influential paper by Arkani-Hamed, Finkbeiner, Slatyer, and Weiner (2009); and the scalar field equation~(2) for Fuzzy Dark Matter in the influential discussion by Hui, Ostriker, Tremaine, and Witten (2017). But there are other possibilities. Maybe the dark matter is in whole or part black holes that were present before the earliest stars. The thought has a long history (e.g., Zel'dovich and Novikov, 1967; Carr and Hawking, 1974) and continues to look interesting (e.g., Carr, K{\"u}hnel, and Sandstad, 2016; Cappelluti, Hasinger, Natarajan, 2022). Or maybe dark matter is something new. 

Do galaxies always contain dark matter? Since most of the dark matter is in the outskirts of a galaxy, possible exceptions would be galaxies that have been tidally stripped, and the dwarfs that might have formed by dissipative settling from tidal streams. Apart from such effects, dark halos are expected to be universal in the standard $\Lambda$CDM cosmology. There are possibly interesting challenges to this prediction. The large S0 galaxy NGC~3115 is in the field, which is unusual because most S0s at this luminosity are in clusters. If it has a dark matter halo with mass typical of its stellar mass then the halo of NGC~3115 must be much less dense than usual, the halo core radius much broader (Cappellari, Romanowsky, Brodie, et al. 2015). The low surface brightness satellites NGC1052-DF2 and NGC1052-DF4 of the elliptical NGC~1052 also look like exceptions. Keim, van Dokkum, Danieli, et al. (2022) argue that ``the dark matter halo masses of these galaxies cannot be much greater than their stellar masses.'' It is too soon to declare a challenge to the $\Lambda$CDM theory from the evidence of galaxies with little dark matter, but the development of the empirical evidence will be worth following. 

What use is dark matter anyway? We could live on a planet in a solar system in a universe that is identical to ours except that the matter is all baryonic in standard forms, with no dark matter. The larger baryon density would result in a lower residual ionization and a larger molecular hydrogen abundance at decoupling, and the onset of galaxy formation would be delayed by the coupling of all matter to the CMB up to redshift $z\sim 1000$. I am not aware of analyses of the effects on the formation of stars and planets in young galaxies, but there are so many stars in our universe that I expect this alternative universe without dark matter would have ample homes for observers. We could live in a universe similar to ours except that the baryon mass fraction is much smaller, though not zero. Gravity would gather dark matter in halos similar to those of our galaxies, but with far fewer baryons. The dissipation of energy by these baryons would be slowed by the smaller baryon density, though aided by the larger residual ionization allowed by the lower baryon density. If this universe continued expanding into the sufficiently remote future then the baryons in a massive dark halo such as the one around the Milky Way would eventually lose enough energy to become dense enough to collapse to stars and planets and observers. These observers would see far fewer galaxies forming stars, but it is difficult to see how that would adversely affect their well-being. In short, our presence does not seem require an anthropic explanation of the dark matter. Maybe its presence  is purely accidental. Maybe it is an anomaly to be resolved.

\subsection{Why Dark Energy?}\label{sec:Lambda} 

We have evidence of detection of Einstein's Cosmological Constant, $\Lambda$, from the BAO signature in the CMB angular distribution; the consistent BAO signature in the galaxy spatial distribution; the supernova redshift-magnitude relation; the comparison of stellar evolution ages and the cosmic expansion time; and the dynamical measurements of the cosmic mean mass density. If these pieces of evidence were seriously wrong the consistency of the $\Lambda$CDM cosmology with these very different ways to probe the universe would be far more improbable than most of us would be willing to consider. The community  conclusion is instead that we have a compelling empirical case for the presence of something that acts as Einstein's $\Lambda$. This is an argument of reasonableness, of course, not a theorem.

In the standard cosmology we flourish not long after the cosmological constant $\Lambda$ and the mean mass density in matter made equal contributions to the expansion rate. This curious, one might say unlikely, coincidence used to be considered a good argument against the presence of the $\Lambda$ term and for the scale-invariant Einstein-de~Sitter model in which the universe is expanding at escape speed whenever we happen to measure it. The argument has been falsified; we must learn to live with $\Lambda$. The anthropic principle accounts for the coincidence, at least broadly, by the argument that we are in a universe in the multiverse in which the absolute value of $\Lambda$ is about as large as is consistent with our existence (Weinberg 1989). Must we leave it at that? The issue is pressing because it proves to be difficult to see another way out of the expectation that the quantum vacuum mass density is quite unacceptably large.

It is worth reviewing the case for reality of the quantum zero-point energy. Consistency of the theory and measurements of binding energies of molecules requires taking account of zero-point energies. The well-tested consistency of energy and active and passive gravitational masses requires that this real zero-point energy of  matter gravitates. The same quantum and gravity physics applies to the electromagnetic field. But the sum of the electromagnetic zero-point energies over laboratory wavelengths amounts to a gravitating mass density far greater than allowed in a relativistic cosmology. Jordan and Pauli (1928) recognized the problem and proposed that the zero-point energy of the electromagnetic field is not real. This was despite the empirical evidence they had of the reality of the zero-point energy of matter fields. How could they, and we, justify distinguishing between zero-point energies based on the same physical theory? 

There also are the positive and negative zero-point energies of all the other fields of particle physics, with prescriptions for ultraviolet truncation, to be added to the contributions to the stress-energy tensor by field binding energies. This looks  challenging to get right, but the sum surely is vastly different from an acceptable mass density in the relativistic cosmology.

Standard physics is independent of the velocity of the observer. If this is true of the quantum vacuum energy then in general relativity its stress-energy tensor must be of the form $g_{\mu\nu}\Lambda_{\rm qm}$, where $\Lambda_{\rm qm}$ is a constant. This is the form of Einstein's cosmological constant. It would be an elegant result except for the woeful difference between $\Lambda_{\rm qm}$ and the empirical $\Lambda$. 

What are we to make of this? Maybe a symmetry principle to be discovered forces the value of the quantum vacuum energy density to vanish, the only natural and reasonable value. Then the cosmological $\Lambda$ would have to be a new parameter of nature. One could instead turn to the anthropic argument discussed in Section~\ref{sec:Anthropic}. Or maybe the value of  $\Lambda$ is decreasing from the large value postulated in the inflation picture, and is approaching its natural value, zero, but slowly enough that we flourish while $\Lambda$ still is slightly different from zero. This is discussed further in Section~\ref{sec:EvolutionPhysics}. These thoughts are awkward. Maybe nature agrees and has something better for us to find.  

\subsection{Why Magnetic Fields; Why not Cosmic Strings?}

The existence of cosmic magnetic fields is clear; it ranks as an anomaly because of the difficulty of accounting for its presence (e.g. Wielebinski and Beck 2005). Cosmic strings and other topological defects are anomalous because they are natural extensions of standard particle physics, yet they are not part of the standard $\Lambda$CDM theory. As as been said of particle theory, ``what is not forbidden is required.'' This hint from a proven theory is not to be lightly disregarded. 

The magnetic field threaded through the Milky Way is made visible by the tendency of interstellar dust particles to be aligned with long axes either perpendicular or parallel to the local magnetic field. The dust absorbs starlight and reradiates the energy at longer wavelengths, preferentially with the electric field of the radiation parallel to the alignment of the dust. The effect is observed in the polarization of starlight that passes through dust clouds and is partially absorbed, and it is observed in the polarization of the radiation reemitted by the dust at longer wavelengths. I recommend looking at the wonderful map of the magnetic field threading the Milky Way from observations of this polarized radiation obtained by the ESA Planck Satellite.\footnote{Click on \url{https://www.esa.int/ESA_Multimedia/Missions/Planck/(result_type)/images} and scroll down.}  At greater distances and on larger scales the evidence of magnetic fields is less direct and an active line of research.

The formation of cosmic magnetic fields might be understood within standard physics applied to what is known of the astrophysics (e.g., Daly and Loeb 1990; Kulsrud and Zweibel 2008; Durrer and  Neronov 2013; Garaldi, Pakmor, and Springel 2021); or maybe we need new physics (e.g., Ratra 1992; Widrow, Ryu, Schleicher, et al. 2012). Or maybe magnetic fields grew out of fossils from the universe before the big bang, whatever that means.

The thought that cosmic strings and other field defects are natural extensions of the standard model for particle physics was persuasive enough to motivate considerable research on how cosmic strings might be observable and might play a role in the formation of cosmic structure (e.g. Kibble 1980; Vilenkin and Shellard 2000). Although  cosmic strings are absent in the present standard cosmology it remains important to look for their effects in cosmic structure, the angular distribution of the CMB, and the gravitational waves produced by cosmic strings. Vachaspati (2021) reviews the present state of this art. Ostriker, Thompson, and Witten (1986) introduced the fascinating idea of magnetized superconducting cosmic strings; maybe they hold the secret to where cosmic magnetic fields came from.  And maybe cosmic strings have something to do with the curiosities in the large-scale distributions of AGNs and rich clusters of galaxies that are discussed in Section~\ref{sec:distributions}. 

\section{Large-Scale Distributions of Radio Galaxies, Quasars, and Clusters of Galaxies}\label{sec:distributions}

Analytic estimates and numerical simulations of cosmic structure formation in the $\Lambda$CDM cosmology show the growth of mass concentrations that are good approximations to observed rich clusters of galaxies. There are discrepancies in the cosmological parameters that best fit cluster counts and best fit the other constraints, but they are small (e.g., Perivolaropoulos and Skara 2022, Table~2). I would count this as a success for the standard cosmology if there were not the curious distributions of clusters and powerful radio galaxies at redshifts $z\lap 0.02$, and the distributions of radio galaxies and quasars at distances approaching the Hubble length. A standard interpretation is that these unusual objects form where the ambient mass density is unusually large. The evidence that there is more to it is reviewed in Section~\ref{sec:LSC}, on the situation in the region around us some 170~Mpc across, and in Section~\ref{CosPrin}, on scales closer to the Hubble length. A summary assessment is presented in Section~\ref{sec:remarks}. 

\subsection{The Extended Local Supercluster}\label{sec:LSC} 

\begin{figure}
\begin{center}
\includegraphics[angle=0,width=3.25in]{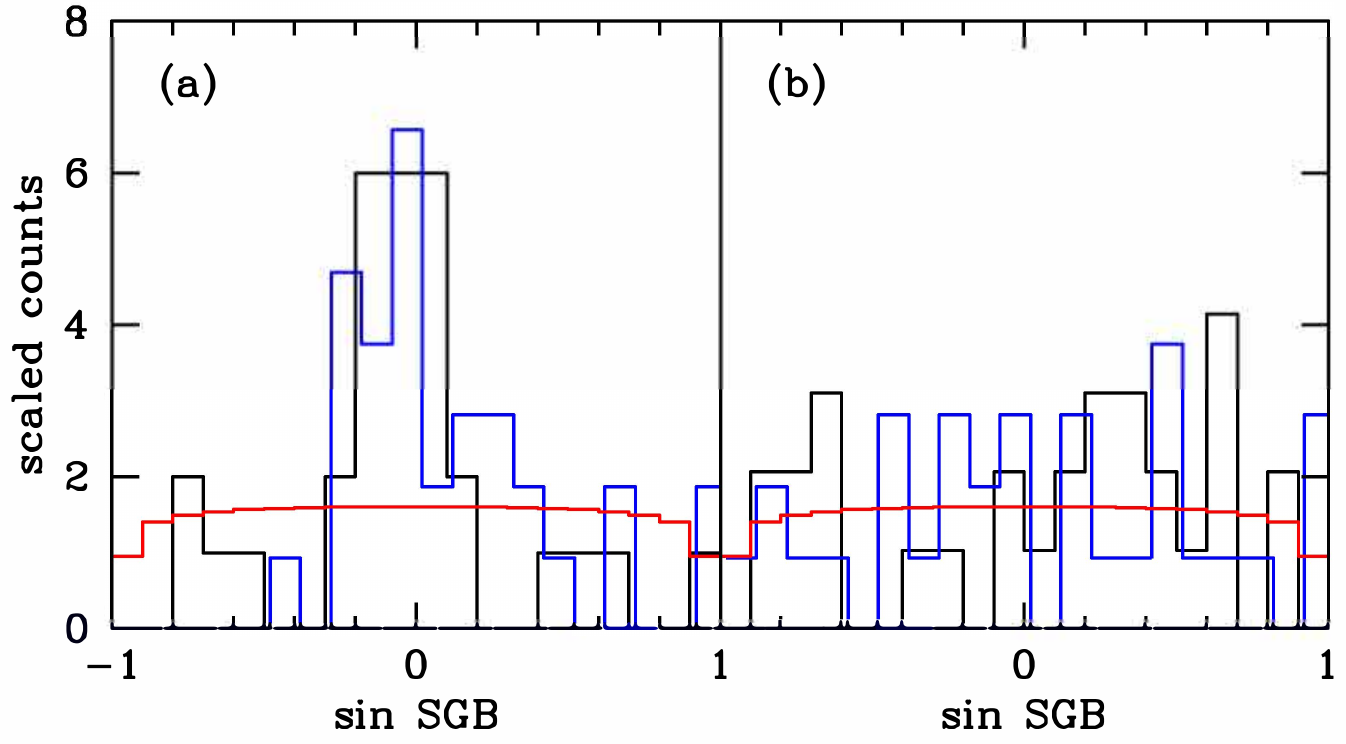} 
\caption{Angular distributions of the sine of the supergalactic latitude of objects at redshifts $z=0.01$ to $0.02$ and galactic latitudes $|b|>10^\circ$. The red histograms are the mean for an isotropic distribution. Panel (a) shows the distributions of the 30 clusters of galaxies in this region plotted in black and the 32 most powerful radio galaxies in blue and shifted slightly to the right. Panel (b) shows the distributions of the 29 galaxies that are most luminous at $60\mu$ plotted in black, and the next 32 most luminous in blue. The counts in the vertical axes are scaled to a common effective area of 30 objects under each histogram. (Figs. adapted from Peebles 2022b.)}\label{fig:LSCf}
\end{center}
\end{figure}

G\'erard de Vaucouleurs' (1953) Local Supercluster is observed as concentrations of relatively nearby galaxies near the great circle across the sky that defines the plane of the Local Supercluster. It includes the concentrations of galaxies in our Local Group and in and around the Virgo Cluster of galaxies at about 18 Mpc distance. The pronounced presence of galaxies near this plane extends to perhaps 30 Mpc. To be considered here is the distributions of objects beyond that distance, in the region at redshifts between $z=0.01$ to 0.02, or distances about 45 to 85~Mpc at Hubble's constant
\beq
H_{\rm o} = 70\hbox{ km s}^{-1}\hbox{ Mpc}^{-1}.
\eeq 
The lower bound on distance removes our special situation close to the plane of the Local Supercluster. The upper bound defines a region about $170$~Mpc across, with the central one eighth of the volume removed. The galaxies are distributed in clumps that look fairly close to uniformly scattered across this region. But the great clusters of galaxies, and the galaxies that are powerful radio sources, tend to be near the extension of the plane of the Local Supercluster. Tully (1986) and Tully, Scaramella, Vettolani, and Zamorani (1992) pointed out this effect for clusters, and Shaver and Pierre (1989) found the same effect for radio galaxies. Shaver (1991) remarked on the key point, the distinct difference from the far weaker concentration of the general population of galaxies to this plane. This interesting difference is not widely advertised, but it it well established and illustrated in Figure~\ref{fig:LSCf}. 

The supergalactic latitude SGB of an object is the angular distance from the great circle defined by the plane of the Local Supercluster. The two panels in Figure~\ref{fig:LSCf} show distributions of the counts of angular positions of objects in equal intervals in sin~SGB, which are equal intervals of solid angle. The data are truncated at galactic latitude $|b|=10^\circ$ to take account of obscuration and confusion near the plane of the Milky Way Galaxy. This reduces the solid angles of the samples, largely at high supergalactic latitudes, SGB close to $\pm 90^\circ$. The effect is seen in the red histograms that show the mean of a random isotropic distribution of points at $|b|>10^\circ$. These red histograms are nearly flat, but suppressed at high supergalactic latitudes by the absence of objects at low galactic latitudes. 

The black histogram in Panel (a) in Figure~\ref{fig:LSCf}  is the distribution in sin~SGB of the 30 clusters of galaxies at $0.01 < z < 0.02$ detected as X-ray sources (from the NASA HEASARC compilation of clusters detected by the X-ray luminosity of the hot intracluster plasma). The clusters are close to the plane of the Local Supercluster as indicated by the peak at low angle SGB. This is to be compared to the red curve expected of an isotropic distribution truncated at low galactic latitude. At lower redshifts the Virgo and Ursa Major clusters, and the rich groups that contain the radio galaxy Centaurus~A and the giant elliptical galaxy IC 3370, are at supergalactic latitudes $-2.4^\circ$, $2.8^\circ$, $-5.2^\circ$ and $-15.1^\circ$, respectively. Again, they are close to the plane.

The blue histogram shifted slightly to the right in Panel (a) is the distribution of the 32 most luminous galaxies at radio wavelengths $\sim 1.1$~GHz (compiled by van Velzen, Falcke, Schellart, Nierstenh{\"o}fer, and Kampert 2012; the data were downloaded from the VizieR Online Data Catalog). This region contains some $10^4$ galaxies with stellar masses comparable to or larger than the Milky Way, luminosities $L\gap L_\ast$, meaning the 32 most powerful radio sources are exceptional galaxies. They tend to be in clusters, but since clusters and radio sources are found and cataloged in very different ways the consistency of their distributions offers a meaningful test of reproducibility of the concentration to the plane indicated by the peaks at low angle from the plane. 

Panel~(b) in Figure~\ref{fig:LSCf} shows the distribution of the galaxies at $0.01<z<0.02$ that are most luminous at $60\mu$ wavelength. This is based on the redshift catalog Saunders, Sutherland, Maddox, et al. (2000) drew from the infrared astronomical satellite sky survey (Neugebauer, Habing, van Duinen, et al., 1984) at wavelengths from 12 to $100\mu$. The data were downloaded from the NASA HEASARC IRASPSCZ catalog, class GALAXY.  I refer to these objects as LIRGs, for Luminous Infrared Galaxies, which seems appropriate though they do not necessarily fit the standard definition. The evidence is that these LIRGs are extraordinarily luminous at $60\mu$ because they are passing through phases of rapid formation of stars that generate the dust that absorbs the starlight and reradiates it as infrared radiation (P{\'e}rez-Torres, Mattila, Alonso-Herrero, Aalto, and Efstathiou 2021 and references therein). The black histogram  in Panel~(b) in Figure~\ref{fig:LSCf} is the distribution in sin~SGB of the 29 most luminous LIRGs. For a check of reproducibility the blue histogram shifted slightly to the right shows the distribution of the 32 next most luminous. Both show no obvious tendency for these galaxies to be close to the plane of the Local Supercluster, or to avoid it. This is a pronounced difference from the distributions of the most powerful radio galaxies in Panel (a), and an illustration of the different distributions of different kinds of galaxies on scales $\sim 100$~Mpc.

\begin{figure}
\begin{center}
\includegraphics[angle=0,width=3.25in]{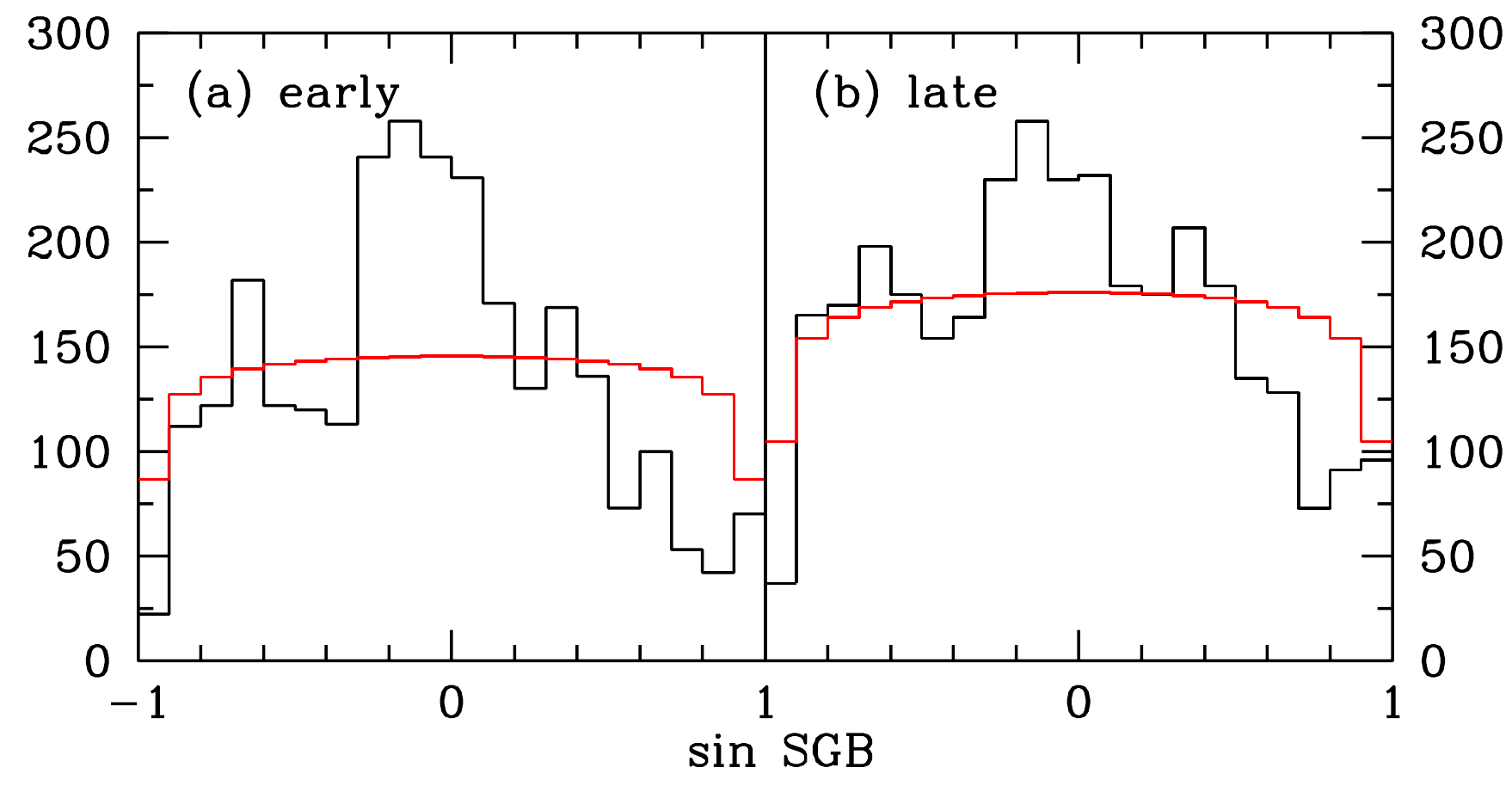} 
\caption{The same as Figure~\ref{fig:LSCf} for the distributions of more numerous  early- and late- type galaxies in the Huchra et al. (2012) catalog.}\label{fig:commongalaxies}
\end{center}
\end{figure}

Figure~\ref{fig:commongalaxies} shows the distribution relative to the plane of the Local Supercluster of a more numerous sample of galaxies (drawn from the Huchra, Macri, Masters, et al. 2012 catalog based on the Skrutskie, Cutri, Stiening, et al. 2006 identifications  of galaxies detected in the Two Micron All Sky Survey, 2MASS). These galaxies are luminous enough to be in the Huchra et al. catalog out to redshift $z\leq 0.02$. Panel (a) shows the distribution of the 2708 early-type galaxies, ellipticals plus S0s (with Huchra morphological types $T\leq 0$), and Panel (b) shows the distribution of the 3276 spiral galaxies (with $1\leq T\leq 9$).\footnote{S0s are mentioned here for completeness, but since S0s are not common among the nearby $L\sim L_\ast$ galaxies they do not figure much in this essay, as noted in footnote~\ref{fn:S0s}. It might be interesting to compare distributions of separate spiral types, but this is not considered here.} 

Since the early-type elliptical and S0 galaxies tend to be in clusters it is not surprising that the distribution in Panel~(a) is peaked at low SGB, with the clusters. But the peak is not as pronounced as in Panel~(a) in Figure~\ref{fig:LSCf}, meaning a greater fraction of these less luminous early-type galaxies are well away from the plane compared to the most powerful radio galaxies.

If considered alone the peak in the distribution of the 3276 spiral galaxies in Panel~(b) in Figure~\ref{fig:commongalaxies} might not seem significant; perhaps it only shows the large fluctuations to be expected in the correlated positions of galaxies, in this case a deficit at positive SGB balanced by an excess at SGB close to zero. But the peaks in the other distributions argue for a real tendency of spiral galaxies to be near this special plane. We also see that the tendency of common spirals to be near this plane is weaker than for common ellipticals plus S0s, which in turn is weaker than for powerful radio galaxies. 

\begin{figure}
\begin{center}
\includegraphics[angle=0,width=3.25in]{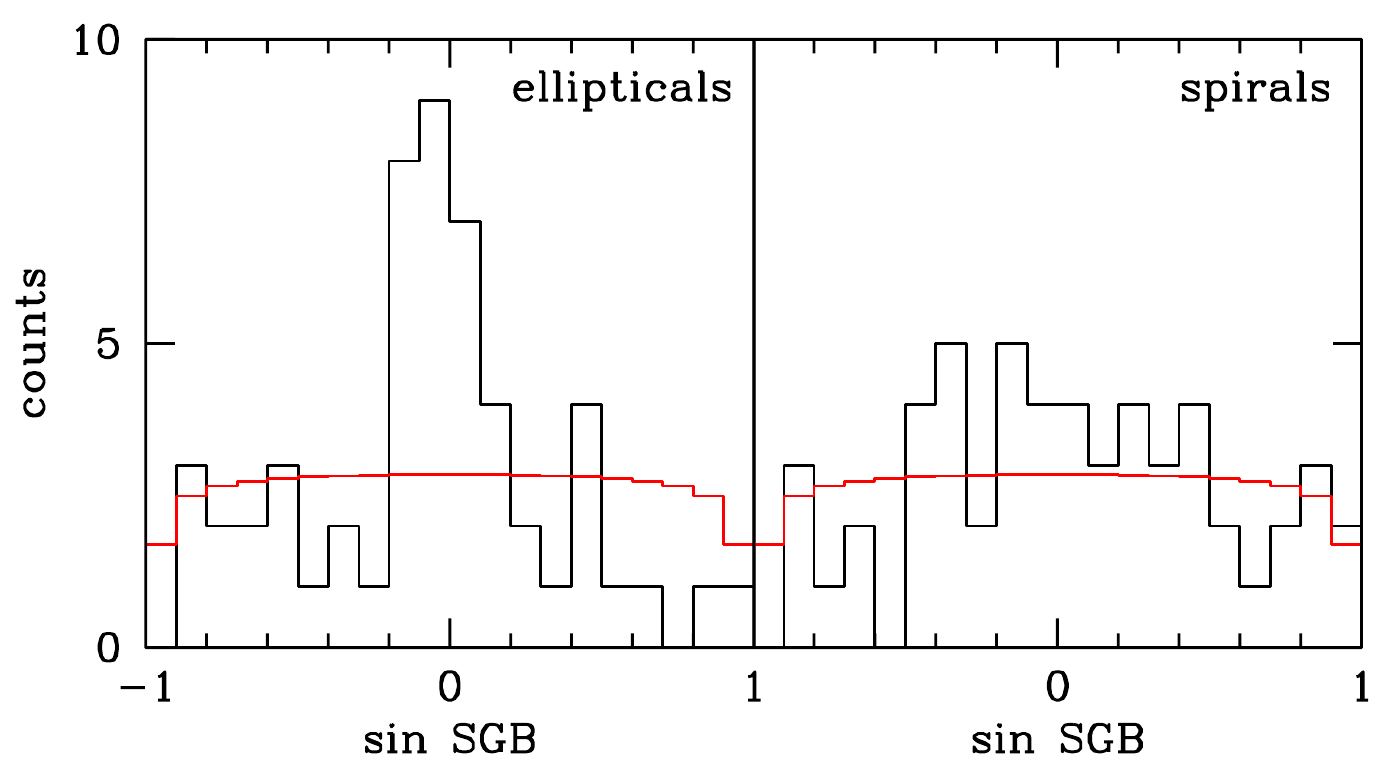} 
\caption{The same as Figures~\ref{fig:LSCf} and~\ref{fig:commongalaxies} for distributions of the most luminous elliptical and spiral galaxies (adapted from Peebles 2022b).}\label{fig:morphologies}
\end{center}
\end{figure}

Figure~\ref{fig:morphologies} shows another aspect, a  comparison of the distributions in sin~SGB of the elliptical and spiral galaxies with the greatest stellar masses (as indicated by the luminosity at $\sim 2\mu$ wavelength, which is considered a useful indicator of the stellar mass). These data also were drawn from the Huchra, et al. (2012) catalog. The 180 most luminous galaxies in the catalog at  $0.01<z<0.02$ and galactic latitudes $|b|>10^\circ$ have absolute magnitudes bounded at apparent magnitude $K_s < 9.5$ at $z=0.02$. Of them, 53 are classified as ellipticals (with Huchra classification $T\leq -5$), 54 are classified spirals ($1\leq T\leq 9$), and the rest are classified S0s and irregulars of various kinds. Figure~\ref{fig:morphologies}  shows the distributions in sin~SGB of the ellipticals and spirals.

It is not surprising that the angular distribution of the 53 most luminous ellipticals is similar to that of the most powerful radio galaxies and the clusters of galaxies shown in Figure~\ref{fig:LSCf}, because radio galaxies tend to be in giant ellipticals that tend to be in clusters. But again the data were obtained in different ways, and the samples are different. The pronounced concentration of the most luminous ellipticals to the plane of the extended plane of the Local Supercluster adds to the evidence of this interesting alignment. 

The 54 most luminous spirals, with stellar masses that are comparable to the 53 most luminous ellipticals, are not noticeably more or less common at low SGB. Their distribution in the right-hand panel in Figure~\ref{fig:morphologies} resembles that of the most luminous galaxies at $60\mu$ in Panel (b) in Figure~\ref{fig:LSCf}. Within the noise there could be a peak in the distribution similar to that of more common spirals in Panel~(b) in Figure\ref{fig:commongalaxies}.

The arrangements of  clusters of galaxies and galaxies of various types relative to the plane of the Local Supercluster are readily studied because we happen to be close to the plane. Other arrangements of objects similar to what is seen within 85~Mpc distance from our special position might include the Pisces-Perseus supercluster (Giovanelli, Haynes, and Chincarini, 1986), the CfA Great Wall (Geller and Huchra 1989), and the Sloan Great Wall (Gott, Juri{\'c}, Schlegel, et al., 2005). They are spread over hundreds of megaparsecs. It would be interesting to know whether radio galaxies and massive ellipticals in the neighborhoods of these more distant configurations are largely confined to a ridge, in the manner of the Local Supercluster, while massive spiral galaxies, and galaxies that are exceptionally luminous in the infrared, are not so particularly concentrated. 

I have not discussed the distribution of quasars at $0.01<z<0.02$ because I am not aware of a quasar catalog with measured redshifts that is suitably close to complete across the sky at this relatively short distance. But we might take it that quasars and radio galaxies are related, and that the concentration of radio galaxies to a plane in a region around us $\sim 170$~Mpc across is an analog of a Large Quasar Group of the kind discussed by Clowes, Harris, Raghunathan, et al. (2013). 

In the sample at $0.01 < z < 0.02$ the similar distributions of giant ellipticals, radio galaxies, and clusters of galaxies on the one  hand, and the different but again similar distributions of giant spirals and luminous infrared galaxies on the other, present us with interesting issues.

\begin{enumerate}[label*=\arabic*.]

\item  Why are the clusters of galaxies largely near a plane at $z<0.02$? If clusters formed where primeval mass density fluctuations were particularly large it would require that upward density fluctuations large enough to grow into clusters are confined to a plane in a region some 170~Mpc across. These large upward density fluctuations cannot be at all common at similar distances from us but not near the plane, because clusters are not at all common there. On the face of it this arrangement looks unlikely in the standard $\Lambda$CDM theory, but it could be checked in pure dark matter simulations.

\item Why are primeval conditions capable of producing galaxies with exceptionally large stellar masses far from the plane as well as near it, but only capable of producing clusters near the plane?

\item Why are the most massive elliptical and spiral galaxies, with comparable stellar masses to judge by the $2\mu$ luminosities, so differently distributed? If the clusters formed where the primeval mass density is exceptionally large, close to this preferred plane, it could account for the concentration of giant ellipticals to this plane. But then why are the giant spirals not more noticeably abundant than average near the plane of the Local Supercluster? Maybe the conditions that favor cluster formation were hostile to the formation of spirals near this plane, perhaps more likely to destroy the spiral arms? But if so why do the spirals not show evidence of avoiding the plane? The more common spirals instead show a modest tendency to be near the plane.

\item What accounts for the concentration of powerful radio galaxies to this plane? Radio galaxies seem to require the presence of a massive compact central object, very likely a black hole. Massive black holes are present in many if not all $L\sim L_\ast$ galaxies, spirals as well as ellipticals, including the many that are closer than 85~Mpc and not close to the plane of the Local Supercluster. These more common black holes seem to cause some galaxies to be radio sources, though seldom at the level of power of the radio galaxies that tend to be near the plane of the extended Local Supercluster. What is special about the massive black holes that are associated with the powerful radio sources that tend to be close to this plane?

\item What are we to make of the contrast between the distributions of galaxies that are exceptionally luminous  at radio wavelengths and those that are exceptionally luminous at $60\mu$? The former tend to be near the plane, the latter not. 

\end{enumerate}

These questions are not widely advertised. A measure of this is the 26 citations to Shaver (1991) in the Astrophysics Data System. One is a self-citation, four are mine, all on Shaver's point, five are on the possible implication of the alignment of radio galaxies for the angular distribution of energetic cosmic rays, and sixteen are on the nature of the space distribution of radio galaxies. Three of these space distribution papers take note of the planar distribution of the relatively nearby radio galaxies and clusters of galaxies. For example, Strauss (1993) remarks on ``the very large planar structures seen in the cluster and radio galaxy distribution by Tully et al. (1992) and Shaver (1991); this discrepancy remains to be explained.''  It certainly is interesting. But I find no discussion of what is to me most interesting, the considerable difference between the space distributions of radio galaxies and clusters of galaxies on the one hand, and the distribution of ordinary large galaxies on the other. The literature on this point might be scant to nonexistent because it is difficult to know what to make of it. That is no excuse, though: we are missing something, which surely is worth investigating.

Figures~\ref{fig:LSCf} to \ref{fig:morphologies} are added illustrations of Shaver's (1991) key point, that different kinds of extragalactic objects can have quite different spatial distributions relative to the plane of the Local Supercluster. Hints to understanding this might be found in the situation on a larger scale that seems analogous, as discussed next. 

\subsection{Anomalies on Large Scales}\label{CosPrin}

To be considered here are apparent anomalies in the distributions and motions of objects on scales approaching the Hubble length. This begins in Section~\ref{sec:CMBDipole} with our motion relative to the reference frame set by the near homogeneous sea of thermal cosmic microwave background radiation, the CMB. The standard interpretation of the CMB dipole anisotropy is that it is the effect of our motion through the radiation. This is tested by computing the local peculiar velocity expected from the gravitational acceleration computed from the observed departures from a homogeneous distribution of objects that seem likely to be useful mass tracers. The results discussed in Section~\ref{sec:localgravity} do not disagree with the idea, but they are not very tight. The test considered in Section~\ref{sec:bulkflows} uses estimates of the mean motion relative to the CMB of objects within a given distance from us, computed from the departures of redshifts from the homogeneous Hubble flow. By some measures this mean motion, the bulk flow, is found to approach zero as the distance is increased, about as expected from the standard $\Lambda$CDM theory. But other measures of the bulk flow that look equally reliable are anomalous. A possibly related problem reviewed in Section~\ref{sec:KinematicDipole} is the predicted dipole anisotropy in the angular distributions of objects that are so far away that, in the standard cosmology, the space distribution likely averages out to homogeneity. The Doppler shifts and aberration of the observed angular positions of these objects caused by our motion relative to the CMB are predicted to produce a dipole anisotropy in the angular distributions of these objects, the kinematic dipole. The measured dipoles in the distribution of quasars and in the distributions of radio galaxies cataloged at several radio frequencies are in about the predicted direction, but the dipole amplitudes are too large, an anomaly. The situation from these considerations is reviewed in Section~\ref{sec:remarks}.

\subsubsection{The CMB Dipole Anisotropy}\label{sec:CMBDipole}

Departures from an exactly homogeneous sea of thermal microwave radiation, the CMB, are usefully represented by the spherical harmonic expansions of the CMB temperature and polarization as functions of position across the sky (eq.~\ref{eq:harmonicexp}). The amplitudes $a_\ell^m$ of the spherical harmonic expansion of the temperature at degree $\ell > 1$ are convincingly demonstrated to be remnants of the decoupling of acoustic oscillations in the plasma-radiation fluid. The dipole amplitude, $\ell = 1$, is much larger than predicted from this effect. The standard idea attributes it to our motion relative to the rest frame defined by the CMB, at velocity  
\beq
{\vec v}_{\rm helio} - {\vec v}_{\rm CMB} = 370\hbox{ km s}^{-1}\hbox{ to } l = 264^\circ,\ b=48^\circ,\label{eq:heliocen_wrt_cmb}
\eeq
in galactic coordinates and the solar system rest frame (Planck Collaboration et al. 2020a). The adjustment for the solar motion relative to the Milky Way Galaxy and the motion of the Galaxy relative to an estimate of the motion of the center of mass of the Local Group, between the Milky Way and M~31, indicates the Local Group of galaxies is moving relative to the sea of radiation at  
\beq
{\vec v}_{\rm Local Group} - {\vec v}_{\rm CMB}= 620\hbox{ km s}^{-1}\hbox{ to } l = 272^\circ,\ b=30^\circ. \label{eq:LG_wrt_cmb}
\eeq
The speed, $\sim 600$~km~s$^{-1}$, is much larger than the relative motions of the galaxies closer than 10~Mpc. A natural interpretation is that we and the nearby galaxies are moving at a near common velocity relative to the CMB, and that this motion is to be associated with the growing departures from an exactly homogeneous mass distribution. A test is discussed next.

\subsubsection{The Peculiar Gravitational Acceleration} \label{sec:localgravity}

In linear perturbation theory and time-orthogonal coordinates chosen to eliminate the decaying mode in a cosmologically flat universe our peculiar motion at coordinate position $\vec r=0$ is predicted to be usefully approximated (in a version of eq.~[\ref{eq:peculiaracceleration}]) as 
\begin{equation}
\vec v = {2\beta G \rho_b\over 3 H_{\rm o}\Omega}\int d^3r~\delta_g(\vec r)~{\vec r\over r^3},\label{eq:dynamic_v}
\end{equation}
where
\begin{equation}
\delta_g(\vec r) = {\delta n(\vec r)\over\langle n\rangle},\quad
\delta_\rho={\delta\rho(\vec r)\over\langle\rho\rangle}\simeq{\delta_g\over b},\quad  \beta \approx {\Omega^{0.55}\over b}\sim 0.4.\label{eq:dynamicparameters}
\end{equation}
The fractional departure from a homogeneous galaxy distribution, $\delta_g(\vec r)$, might be represented as a sum of Dirac delta functions minus the mean, or the result of smoothing of galaxy counts through a window. The  mass density contrast $\delta_\rho$ is written in the simple linear bias model for the relative distributions of galaxies and mass. We have a measure of the bias parameter, $b\sim 1.2$, and the mass density parameter, $\Omega=0.31$, from the fit of theory to observations of the patterns in the distributions of galaxies and the CMB (as discussed in Planck Collaboration et al. 2020b). The evidence from these data is that galaxies are reasonably good mass tracers. 

Having adopted general relativity in the standard cosmology we must accept that the mass outside the region we can observe in principle (that is, outside the particle horizon subsequent to inflation, or whatever saved us from the singularity of the standard model) can affect predicted observable peculiar motions (as discussed in Sec.~\ref{sec:theacausaluiverse}). There is evidence that the mass distribution that can be observed is a useful approximation, however. Erdo{\v{g}}du, Huchra, Lahav, et al. (2006) numerically evaluated the integral in equation~(\ref{eq:dynamic_v}) using a version of the Huchra, Macri, Masters, et al. (2012) galaxy redshift catalog. (This valuable catalog was used in the studies of the extended Local Supercluster in Sec.~\ref{sec:LSC}.) Erdo{\v{g}}du et al. found that the predicted velocity of the Local Group relative to the CMB seems to converge at about 100~Mpc distance from the Local Group. The integral computed to this distance indicates the motion of the Local Group is toward $l\sim 265^\circ$, $b\sim 38^\circ$, some $10^\circ$ from the CMB dipole. This is tolerably consistent considering the uncertainties in the use of galaxies as mass tracers. The integral computed to this distance agrees with the observed speed if the combination $\beta$ of bias parameter and mass density parameter in equation~(\ref{eq:dynamicparameters}) is $\beta = 0.40\pm 0.09$. This is consistent with the value derived from the CMB anisotropy spectrum (Planck Collaboration et al. 2020). At greater cutoff distances the computed velocity moves away from the CMB direction, a likely consequence of large effects of small systematic errors in the mass distribution on larger scales.

This test is important but not yet very precise. A better application requires a catalog of positions of useful mass tacers to greater distances with tighter controls on completeness and systematic errors in distances and positions across the sky, a challenging task. And we must live with the unknowable situation outside the Hubble length.

\subsubsection{Cosmic Bulk Flows}\label{sec:bulkflows}

This probe requires measurements of the radial components $v_p$ of peculiar velocities of objects relative to the general expansion of the universe, 
\beq
v_p = cz - H_{\rm o}r. \label{eq:peculiarvelocity}
\eeq
To order $v/c$ the radial velocity derived from the measured redshift $z$ of an object is $v =  cz$, $H_{\rm o}r$ is the speed of cosmological recession at the physical distance $r$ of the object, and the difference is the radial peculiar velocity $v_p$. In a suitably large sample of objects that are moving relative to us at mean velocity $\vec v_{\rm obs}$ the measured  $v_p$ of the objects are expected to vary across the sky as $v_{\rm obs}\cos\alpha$, where $\alpha$ is the angle between the direction to the object and the direction of the mean velocity $\vec v_{\rm obs}$. It is conventional to define the bulk velocity $\vec v_{\rm sample}$ of the sample to be the mean velocity referred to the rest frame in which the CMB has no dipole anisotropy. (This ignores the small intrinsic dipole remnant from the decoupling of baryonic matter and the CMB.) In the standard model $\vec v_{\rm sample}$ is predicted to converge to zero in a catalog that reaches distances large enough to be a fair sample of the statistically homogeneous universe. 


Boruah, Hudson, and Lavaux (2020) measured peculiar velocities of galaxies closer than $\sim 60$~Mpc from distances to supernovae of type Ia and distances to spiral galaxies based on the Tully-Fisher relation. The Boruah et al. mean of the peculiar velocities relative to the CMB is 
\beq
{\vec v}_{\rm galaxies} - {\vec v}_{\rm CMB}= 252 \pm 11\hbox{ km s}^{-1}\hbox{ to } l = 293^\circ,\ b=14^\circ  (\pm 5^\circ).\label{eq:Hudson}
\eeq
This can be compared to the earlier Ma and Scott (2013) measurement on a similar scale, 
\beq
{\vec v}_{\rm galaxies} - {\vec v}_{\rm CMB}\sim 290\pm 10 \hbox{ km s}^{-1}\hbox{ to } l = 280\pm 8^\circ,\ b=5^\circ \pm 6^\circ.\label{eq:Scott}
\eeq
from fundamental plane distances to early type galaxies, Type Ia supernovae, and the Tully-Fisher relation for late-type galaxies. (The speed is my estimate from the four results in Ma and Scott Table 2.) Similar results are reported in quite a few analyses. Watkins, Feldman, and Hudson (2009) found bulk flow direction consistent with equations~(\ref{eq:Hudson} and~(\ref{eq:Scott}) and speed 150~km~s$^{-1}$ larger, but that is just twice the estimated uncertainty. Tighter and consistent results are reported by Turnbull, Hudson, Feldman, et al. (2012), who used Type~Ia supernova distances; Hong, Springob, Staveley-Smith, et al. (2014), who used Tully-Fisher distances to spiral galaxies; and Scrimgeour, Davis, Blake, et al. (2016), who used the fundamental plane relation to get distances to early-type galaxies. This bulk velocity seems to be securely established.

The bulk velocity of the sample of objects closer than about 60~Mpc need not be in the same direction as the peculiar velocity of the Local Group, but one might expect it to be fairly close, as it is in these measurements. The speed of a sample relative to the CMB, the bulk velocity, is expected to be smaller when averaged over larger scales, because the mass distribution is assumed to average to homogeneity in sufficiently large volumes. The speeds found from these analyses are roughly half that of the Local Group, and Figure 10 in Boruah et al. indicates the speed is consistent with the probability distribution in the average over a region of this size computed in linear perturbation theory in the $\Lambda$CDM cosmology.  

In a remarkable advance the Planck Collaboration et al. (2014) reported a measurement of the mean motion of clusters of galaxies relative to the CMB based on the kinematic  kSZ effect (Sunyaev and Zel'dovich 1980). This is the Doppler shift of the CMB scattered by electrons in the intracluster plasma of a cluster of galaxies that is moving relative to the CMB. The conclusion from detections of this kSZ effect is that the average speed of the clusters relative to the CMB is compatible with zero and the speed is less than about $260\hbox{ km s}^{-1}$ at the 95\% confidence level.This is in a sample of clusters with redshifts ranging around $z\sim 0.2$. 

The detection of the kSZ effect in the plasma in a cluster of galaxies offers a direct measure of the effect of the motion of the cluster relative to the CMB. The Planck Collaboration observations agree with the expectation that the motion of the clusters averaged over scales approaching the Hubble length is small, continuing the trend from the speed of the Local Group, $\sim 600$~km~$s^{-1}$, to the mean for the galaxies out to 60~Mpc at half that speed, to the still smaller mean speed expected of the still more extended sample of clusters of galaxies. 

The motions of clusters of galaxies based on measurements of distance rather than the kSZ effect are more complicated. Lauer and Postman (1994) used distances to Abell clusters  (Abell 1958; Abell, Corwin, Olowin, 1989) based on apparent magnitudes of the brightest cluster members, measured in an aperture of fixed metric size and corrected for the redshift of the spectrum and extinction in the Miky Way. They found that the mean peculiar velocity of the Abell clusters relative to the CMB is
\beq
{\vec v}_{\rm Abell} - {\vec v}_{\rm CMB} \sim 689\hbox{ km s}^{-1}\hbox{ to } l = 343^\circ ,\ b = 52^\circ.\label{eq:LP}
\eeq
The direction might be taken to be roughly similar to that of the galaxies (eq.~[\ref{eq:Hudson}]), but the large speed is anomalous. This result remains unexplained. The Migkas, Pacaud, Schellenberger et al. (2021) measurement of the cluster bulk flow used scaling relations among properties of the plasma in clusters rather than properties of the cluster galaxies. The smallest scatter among their measurements is found in the relations between the distance-independent intracluster plasma temperature and the distance-dependent cluster X-ray luminosity,  and between the plasma temperature and the distance-dependent integrated Comptonization parameter. The clusters in their sample are at redshifts $z\sim 0.01$ to 0.55, with 50\%\ within $z=0.05$ to 0.18. They found that in a region of the sky centered around $l\sim 276^\circ$, $b \sim -16^\circ$ their cluster peculiar velocities referred to the CMB rest frame are systematically negative, with a suggestion of positive peculiar motions in the opposite part of the sky. It can be interpreted as the effect of a mean bulk flow of the cluster sample at velocity
\beq
{\vec v}_{\rm cluster} - {\vec v}_{\rm CMB} \sim 900\hbox{ km s}^{-1}\hbox{ to } l \sim 88^\circ,\ b \sim 16^\circ.\label{eq:Migkasetal}
\eeq
Again, the speed is anomalously large. The direction disagrees with the Lauer and Postman result (eq.~\ref{eq:LP}), and it is close to opposite to the motions of the Local Group and the bulk flow of the galaxies within distance $\sim 60$~Mpc (eqs.~\ref{eq:LG_wrt_cmb}, \ref{eq:Hudson}).

The situation is interesting. The Abell cluster catalog was compiled by hand, so there may be sampling inhomogeneity, but that need not seriously affect the mean motion of the observed clusters. The Lauer and Postman (1984), and  Migkas, et al. (2021), bulk flows are seriously different, maybe because their distances are based on different cluster scaling relations. It also might have something to do with the odd distributions of clusters and radio galaxies closer than 85~Mpc (Sec.~\ref{sec:LSC}), and the odd larger-scale distributions of radio galaxies and quasars to be considered next. 

\subsubsection{The Kinematic Dipole}\label{sec:KinematicDipole} 

This cosmological test requires a catalog of objects that are far enough away that we can assume their mean spatial distribution is adequately close to the homogeneity and isotropy of the cosmological principle. It of course requires that the efficiency of detection of objects is adequately close to uniform across the sky. In these conditions our motion relative to the mean of this sample is expected to produce a dipole anisotropy in the counts of objects, the results of the Doppler shifts of apparent magnitudes and the aberration of angular positions. 

Ellis and Baldwin (1984) pointed out that this consideration provides us with a cosmological test: the dipole anisotropy of counts of objects across the sky is predicted to be, to order $v/c$,
\beq
\delta N/N = [2+x(1+\alpha)](v/c)\cos\theta, \label{eq:EllisBaldwin}
\eeq
where
\beq
 S\propto \nu^{-\alpha}, \ N(>S) \propto S^{-x}. \label{eq:EBparameters}
\eeq
The parameter $\alpha$ is a measure of the typical spectrum of an object, $x$ is a measure of the variation of counts of objects with limiting flux density $S$, and $v$ is the heliocentric velocity relative to the mean of the sample of distant objects.

The physics of the Ellis and Baldwin kinematic effect is well established in other contexts. The Compton-Getting effect is the dipole anisotropy of energetic cosmic rays seen by an observer moving through an isotropic sea of cosmic rays. The same effect accounts for the thermal spectrum of the CMB detected in a given direction by an observer moving relative to the CMB rest frame (Peebles and Wilkinson 1968). Kaiser (1987) remarked that this apparent dipole in the mass distribution produces an apparent contribution to the computation of our cosmic gravitational acceleration. Bahr-Kalus, Bertacca, Verde, and Heavens (2021) review current studies of this ``rocket effect." 

The evidence to be reviewed here is that the dipole anisotropy in the distribution of objects at distances comparable to the Hubble length is about in the direction expected from the kinematic effect if the dipole anisotropy in the CMB is due to our motion relative to the rest frame defined by the mean mass distribution, but the dipole amplitude is at least twice the prediction. This anomaly is about as well established as the Hubble Tension, yet the  literature on the kinematic effect is much smaller than the 344 papers with the phrase ``Hubble Tension'' in the abstract in the SAO/NASA Astrophysics Data System. (I expect the difference is an inevitable consequence of the way we behave.) To illustrate this difference I offer my attempt at a close to complete literature on the kinematic effect (with apology for overlooked publications). 

Baleisis, Lahav, Loan, and Wall (1998) considered the possible detection of the kinematic  effect in the dipole anisotropy of radio galaxies. Since these objects are detectible past redshift $z=1$ their distribution probes scales large enough that one might hope the clustering of these objects averages out to the wanted uniformity to be seen broken by the kinematic dipole. Scharf, Jahoda, Treyer, et al. (2000) considered detections of the kinematic effect in the angular distributions of X-ray, AGNs, and clusters of galaxies. These are the earliest empirical studies of the effect I have found. Blake and Wall (2002) had the NRAO VLA Sky Survey of radio sources (NVSS; Condon, Cotton, Greisen, et al., 1998).  Their dipole estimate is consistent with the direction and amplitude expected from the dipole anisotropy of the CMB. Later studies of the NVSS catalog confirmed the direction, but found larger than expected dipole amplitudes in this catalog at various flux density cuts (Singal, 2011; Gibelyou and Huterer, 2012; Tiwari, Kothari, Naskar, Nadkarni-Ghosh, and Jain 2015; Secrest, von Hausegger, Rameez, Mohayaee, and Sarkar 2022). Rubart and Schwarz (2013) had a partial check by combining NVSS with the Westerbork Northern Sky Survey (WENSS; de Bruyn, Miley, Rengelink, et al., 2000). Colin, Mohayaee, Rameez and Sarkar (2017) combined NVSS with the Sydney University Molonglo Sky Survey (SUMSS; Mauch, Murphy, Buttery, et al., 2003). Both papers reported the anomaly. Bengaly, Maartens and Santos (2018) reported consistency of separate measurements of the dipole anisotropies derived from the NVSS (USA) and TGSS-ADR1 (India; Intema, Jagannathan, Mooley, and Frail, 2017) catalogs, and  Siewert, Schmidt-Rubart, and Schwarz (2021) reported separate measurements of the NVSS, TGSS-ADR1, WENSS (The Netherlands) and SUMSS (Australia) catalogs. Since each of these four catalog might be affected by its own systematic errors the reasonable consistency of the separately analyzed catalogs is a valuable check of reliability. The dipole directions from these four independently obtained samples average to about 
\beq
l\sim 240^\circ,\ b\sim 30^\circ\hbox{ for radio galaxies.}\label{eq:radiodipole}
\eeq
This is not far from the direction of the heliocentric CMB dipole (eq.~[\ref{eq:heliocen_wrt_cmb}]), as would be expected from the kinematic effect. The radio source dipole amplitudes from the different samples are considerably different, maybe in part  because the amplitude depends on frequency, but all are roughly a factor of 4 times the amplitude expected from our motion relative to the CMB rest frame. Siewart et al. conclude that the dipoles ``exceed the expectations derived from the CMB dipole, which cannot strictly be explained by a kinematic dipole alone.'' 

Darling (2022) reports a different situation from the analysis of the dipole anisotropy of radio sources in the more recent VLASS (USA; Lacy, Baum, Chandler, et al., 2020) and RACS (Australia; McConnell, Hale, Lenc, et al., 2020) surveys. The dipole interpreted as the kinematic effect indicates heliocentric velocity
\beq
{\vec v}_{\rm helio} - {\vec v}_{\rm radio}  \sim 330\pm 130\hbox{ km s}^{-1}\hbox{ to } l \sim 270\pm 55^\circ,\ b \sim 56\pm 25^\circ.\label{eq:Darling}
\eeq
This agrees with the heliocentric velocity, direction and speed, relative to the CMB (eq.~[\ref{eq:heliocen_wrt_cmb}]).

Darling estimates that the ``most permissive'' analysis allows effective velocity 740~km~s$^{-1}$ at three standard deviations. Siewert et al. (2021) estimate that at that speed the kinematic would correspond to dipole amplitude $d\sim 0.01$. But this is well below most of the Siewert et al. radio source dipoles. It is a cautionary example of  the difficulty of establishing this important measurement. 

We have a check on this  situation from the angular distribution of the roughly one million objects selected by their mid-infrared colors to be quasars detected by the Wide-field Infrared Survey Explorer (WISE; Wright, Eisenhardt, Mainzer, et al. 2010). In independently selected catalogs, Secrest, von Hausegger, Rameez, et al. (2021, 2022) and Singal (2021) found dipole anisotropies pointing to (in my estimate of Singal's central value) 
\begin{align}
& l\sim 210^\circ,\ b\sim 45^\circ,\ v\sim 1700\hbox{ km s}^{-1},\hbox{ Singal (2021)},\nonumber\\
&l= 238^\circ,\ b=31^\circ, v \sim 750\hbox{ km s}^{-1},\hbox{ Secrest et al. (2022)}. \label{eq:kinematicdipole}
\end{align}
The directions of these two quasar dipoles are reasonably similar to the directions of the radio dipole (eq.~[\ref{eq:radiodipole}]), and perhaps not unreasonably far from the CMB dipole,  $l = 264^\circ,\ b=48^\circ$ (eq.~[\ref{eq:heliocen_wrt_cmb}]). But with the authors' estimates of the parameters for spectrum and counts in equation~(\ref{eq:EBparameters}) the dipole amplitudes are at least twice that expected from the kinematic dipole (eq.~\ref{eq:EllisBaldwin}) at the velocity expected from the CMB dipole. Darling's (2022) new radio galaxy dipole amplitude puts the three standard deviation upper limit of 750~km~s$^{-1}$ on the effective velocity from the kinematic amplitude, consistent with Secrest et al. but still well below Singal. 

Quasars and radio galaxies are related, but the data on the two were obtained and reduced by quite different methods, and each have been analyzed by two or more independent groups with consistent results that seem to make a consistent case for an anomaly. But the result in equation~(\ref{eq:Darling}) from independent radio data and analysis is an important reminder of the hazards of systematic errors in these measurements. I conclude that the present weight of the evidence from the other measures of the radio dipole and the WISE quasar dipole (eqs.~\ref{eq:radiodipole} and \ref{eq:kinematicdipole}) is that there is an anomalously large dipole common to distant radio galaxies and quasars, but the case is not yet persuasive. 

Several other points are to be noted. First, the general conclusion has been that, within the standard cosmology and a reasonable degree of biasing in the positions of quasars and radio galaxies relative the mass, intrinsic inhomogeneity is not a likely explanation for the anomalous dipole anisotropies of these distant objects  (e.g., Rubart, Bacon, and Schwarz 2014; Tiwari and Nusser 2016; Colin et al. 2017; Dom{\`e}nech, Mohayaee, Patil, and Sarkar 2022; and Murray 2022, who considered the effect of gravitational lensing). This might be checked by the cross-correlation between the surface mass density indicated by CMB lensing with the positions of distant radio galaxies (Robertson, Alonso, Harnois-D{\'e}raps, et al., 2021), but I have not found a discussion of the test. A simple order-of-magnitude argument is that if quasars and radio galaxies were useful mass tracers on scales approaching the Hubble length then the observed dipole anisotropies $\delta N/N\sim 0.02$ would have produced a bulk flow on the order of 2\% of the speed of light, which is absurd.

Second, we have a separate reason to question whether these objects are useful mass tracers, from the curious distributions of objects that contain AGNs closer than 85~Mpc (Sec.~\ref{sec:LSC}). Following this line of thought we might expect that the angular positions of clusters of galaxies at redshifts close to unity have an anomalous dipole anisotropy, perhaps similar to that of quasars and radio sources, though of course not similar to the mass distribution. Perhaps this can be checked with available data. If ordinary $L\sim L_\ast$ galaxies are useful mass tracers, as evidenced by the cosmological tests, then a really deep galaxy catalog would be expected to have dipole direction and amplitude consistent with the kinematic dipole (eq.~\ref{eq:EllisBaldwin}) indicated by our motion defined by CMB dipole. It may be possible to check this.
 
Third, one is tempted to ask whether the dipole anisotropy in the distributions of distant radio sources and quasars, or the Migkas et al. (2021) cluster bulk flow, are somehow related to the plane of the Local Supercluster. There is no indication of that in the directions  they define, supergalactic latitudes $SGB\sim -50^\circ$ and $SGB\sim +50^\circ$. But we do see common evidence of anomalous distributions of objects that contain AGNs.

Fourth, Migkas (private communication 2022) points out that our heliocentric velocity relative to the cluster rest frame found by Migkas et al. (2021, eq.~[\ref{eq:Migkasetal}]) is 
\beq
{\vec v}_{\rm helio} - {\vec v}_{\rm clusters}  \sim 1100\hbox{ km s}^{-1}\hbox{ to } l \sim 280^\circ,\ b \sim 5^\circ.\label{eq:Kostas}
\eeq
This is not far from our effective heliocentric velocities relative to the radio sources in equation~(\ref{eq:radiodipole}) and the quasars in equation~(\ref{eq:kinematicdipole}). Maybe it is only a coincidence, but it is not to be ignored in this confusing situation. 

\subsection{Summary Remarks} \label{sec:remarks}

The association of the CMB dipole anisotropy with the growing mode of the departure from an exactly homogeneous universe is tested by the check of consistency with the prediction from the peculiar gravitational acceleration. There is room for an anomaly because the integral  in equation~(\ref{eq:dynamic_v}) seems to converge at $\sim 100$~Mpc distance while the evidence is that the bulk flow of the galaxies on about the same scale is some 250~km~s$^{-1}$ (eq. [\ref{eq:Hudson}]). But assessing the significance of the discrepancy is difficult because the computation of the integral in equation~(\ref{eq:dynamic_v}) is sensitive to small errors in the large-scale galaxy distribution.

The evidence is that the bulk flow of the galaxies relative to the CMB decreases with increasing sample size about as expected from standard ideas, from about 600~km~$^{-1}$ in the average over distances of a few megaparsecs, to about 250~km~$^{-1}$ averaged out to distances $\sim 60$~Mpc. Detection of the kinetic SZ effect on the intracluster plasma suggests an even smaller bulk flow of clusters of galaxies at distances distributed around $\sim 600$~Mpc. Better checks of the convergence of the galaxy bulk flow and the predicted local peculiar gravitational acceleration from the integral over the mass distribution requires improved  measures of the space distributions of galaxies and mass. Perhaps it will come from the Euclid mission ``to capture signatures of the expansion rate of the Universe and the growth of cosmic structures'' (Percival, Balogh, Bond, et al. 2019). Nadolny, Durrer, Kunz, and Padmanabhan (2021) present a worked example of how this might go.

There are interesting anomalies in the space distributions of radio galaxies, quasars, and clusters of galaxies. The evidence reviewed in Section~\ref{sec:LSC} (and shown in Figs. \ref{fig:LSCf} and  \ref{fig:morphologies}) is that we are in a region some 170~Mpc across in which the most luminous early-type galaxies, powerful radio galaxies, and clusters of galaxies tend to be near the extended plane of the Local Supercluster. In contrast to this the spirals with stellar masses as great as the most massive ellipticals, to judge by the luminosities at $2\mu$, are not noticeably correlated with this plane. The same is true of the most luminous galaxies at $60\mu$, and the much more abundant $L\sim L_\ast$ galaxies that are useful mass tracers.

A possibly related anomaly is found in measurements of the bulk velocities of clusters of galaxies (Lauer  and Postman 1994; Migkas, et al. 2021). Both are based on cluster properties: first-ranked galaxy luminosity or cluster plasma scaling relations. If the anomaly from cluster distance measures persist, and the mean motion of clusters of galaxies relative to the CMB measured by the kSZ effect continues to show a small bulk velocity, then we will be forced to the conclusion that scaling relations of cluster properties are not universal. This is not as extreme at it might at first seem, for recall the anomalous distribution of the nearer clusters, in the region 170~Mpc across. 

Yer another possibly related anomaly is the dipole anisotropies in the angular distributions of presumably distant and on average uniformly distributed radio galaxies and quasars. All recent analyses agree that the dipole is about in the direction expected from the kinematic effect of our motion relative to the CMB. Though not all recent analyses agree on the amplitude, the weight of the evidence is that the radio galaxy and quasar dipole amplitudes are anomalously large. 

We are led to the thought that the properties of large groups and clusters of galaxies, radio galaxies, and quasars defined by colors, were enabled by something that has had a subdominant correlation with the mass distribution  everywhere except in rare situations such as the neighborhood of the extended Local Supercluster. This would be a departure from the Gaussian adiabatic initial conditions of the standard cosmology. Thoughts turn to cosmic strings, or primeval isocurvature fluctuations, or black holes left from some earlier epoch. Or, as I have remarked elsewhere in this essay, something completely different. 

Also to be considered is the indication that positions of physically related quasars are spread over greater lengths than could have grown out of the Gaussian near scale-invariant adiabatic initial conditions assumed in the standard $\Lambda$CDM cosmology (e.g., Clowes, et al. 2014). If these Large Quasar Groups are physically real associations then, under the assumptions of the standard theory, they are associations among objects that have never been causally connected. This is a familiar situation, of course. The Gaussian initial conditions of the standard cosmology also are acausal in the standard model (Sec.~\ref{sec:theacausaluiverse}). 

Some authors conclude that the anomalously large dipole anisotropy of distant quasars and radio galaxies, and the Large Quasar Groups, if physically real, violate the  cosmological principle (e.g., Secrest, von Hausegger, Rameez, et al. 2022). This depends on the definition of this principle, of course. It does not violate the definition explained in Section~(\ref{sec:definition}), but it likely violates the assumption of Gaussian near scale-invariant and adiabatic initial conditions that has served so well for many other cosmological tests. 

\begin{figure}
\begin{center}
\includegraphics[angle=0,width=3.7in]{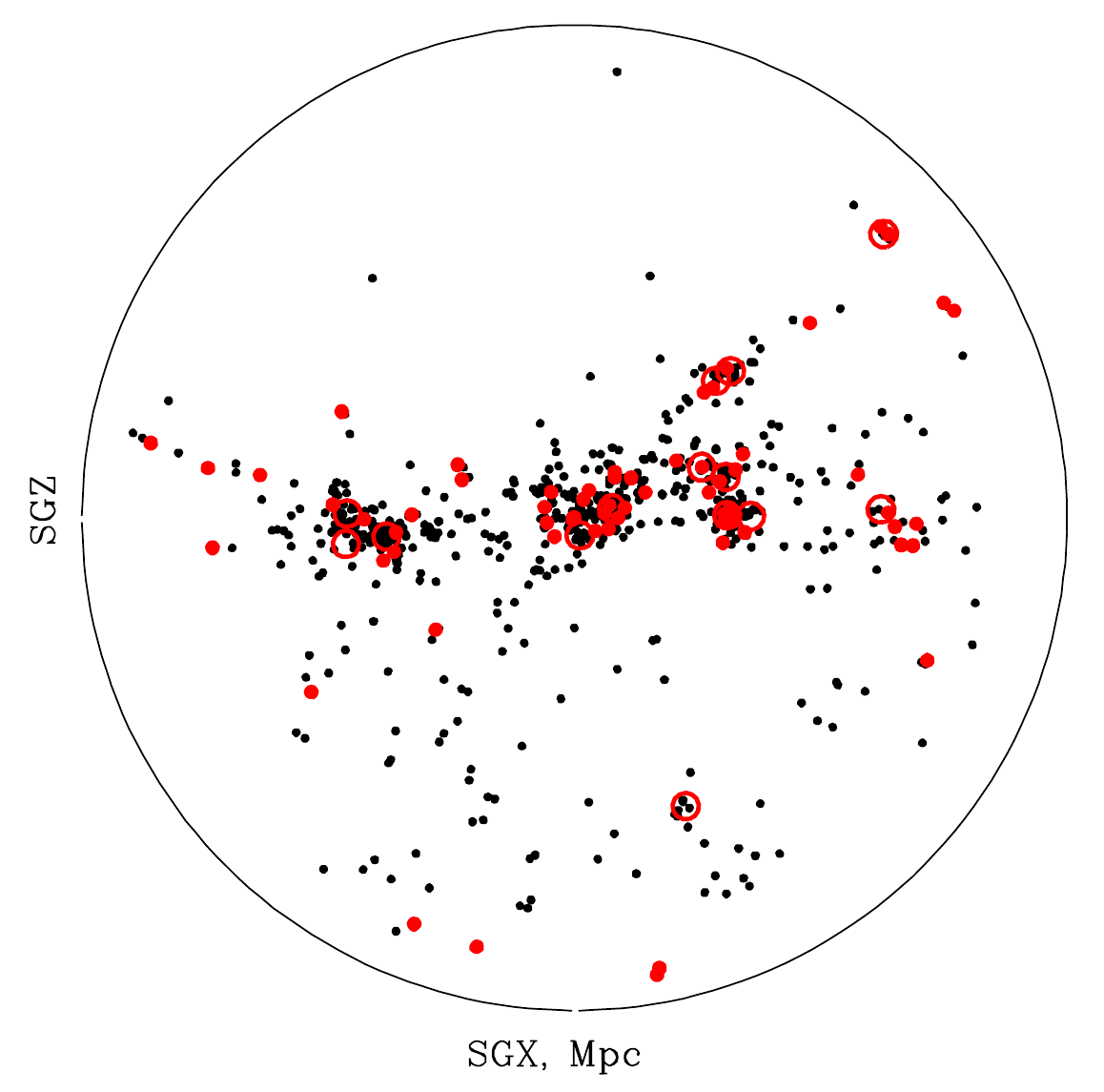} 
\caption{Distribution of galaxies closer than 9~Mpc. The galaxies more luminous than $M_B = -20$ are marked by the large red circles. The smaller  red circles show the positions of the less luminous galaxies with absolute magnitudes $-20 < M_B < -17$. The still smaller black circles show the dwarfs with known distances at $M_B > -17$.}\label{fig:LocalVol}
\end{center}
\end{figure}

\section{The Local Void}\label{sec:localvoid} 

Figure~\ref{fig:LocalVol} shows the distribution of galaxies closer than $D=9$~Mpc, plotted in the supergalactic coordinates discussed in Section~(\ref{sec:LSC}). (The data are from Karachentsev, Karachentseva, Huchtmeier, Makarov, 2004, updated in the NASA HEASARC Updated Nearby Galaxy Catalog. A distance cutoff at $D=8$~Mpc eliminates the interesting dwarf galaxy at the top of the figure; a cutoff at 10~Mpc adds several dwarfs in the low density area to the upper right that are not very close to the two interesting dwarfs that are well within the Local Void.) 

The Local Supercluster is the  concentration of galaxies running through the center of the figure. The distributions of radio galaxies and clusters of galaxies relative to the plane of the Local Supercluster, at ten times the distance sampled in Figure~\ref{fig:LocalVol}, are illustrated in Figures~\ref{fig:LSCf} and~\ref{fig:morphologies}

The open red circles in Figure~\ref{fig:LocalVol} show positions of the 14 most luminous galaxies, absolute magnitudes $M_B<-20$. Galaxies that are more luminous than any but the brightest in this region are exceedingly rare. The 65 smaller filled red circles mark positions of galaxies with absolute magnitudes in the range $-20 < M_B < -17$, which spans a factor of 16 in luminosity. The positions of the 718 less luminous of the galaxies with useful distance estimates are marked as the still smaller filled black circles. It is expected that many more dwarfs will be added to this sample.

The upper left region in Figure~\ref{fig:LocalVol} is part of the Local Void, a strikingly empty region. Only two of the known $\sim 800$ galaxies within 9~Mpc occupy about a quarter of the volume in this part of the Local Void. This amounts to a space density of galaxies in the near empty region at about one percent of the mean within the full $R<9$~Mpc volume. A common estimate from numerical simulations of structure formation in the $\Lambda$CDM theory is that in low density regions the mean mass density bottoms out at roughly 10\% of the cosmic mean. A recent example is presented in Cautun, Cai, and Frenk (2016). Peebles (2001) argued that this low density of detected galaxies seems distinctly odd. Tikhonov and Klypin (2009) concluded from their numerical simulations that  ``The emptiness of voids [is] yet another overabundance problem for the cold dark matter model.'' But Tinker and Conroy (2009) pointed out that the low density of galaxies in the Local Void is consistent with the predicted mass distribution if the most numerous lowest mass dark matter halos contain very few luminous stars. The Tinker and Conroy application of this idea using the halo occupation distribution model applied to high resolution pure dark matter simulations produces acceptably empty voids. This is progress but it is a prescription, not a prediction. It might be tested by other information. 

Let us begin with the two known dwarf galaxies in the nearest part of the Local Void. The dwarf galaxy at the top of the figure, ZOA~J1952+1428, was discovered in a blind survey for HI emission by the Arecibo Zone of Avoidance Survey (McIntyre, Minchin, Momjian, et al. 2011). The other well isolated dwarf galaxy lower down and to the left in Figure~\ref{fig:LocalVol} is KK~246, also known as ESO 461-036. 

Karachentsev, Dolphin, Tully, Sharina, et al. (2006) present HST images of KK~246 among other nearby galaxies. The optical image of KK~246 looks similar to other dwarfs (to my untrained eye) that are not so extremely isolated. Kreckel, Peebles, van Gorkom, van de Weygaert, and van der Hulst (2011) present maps of the extended atomic hydrogen envelope around KK~246. The long axis of the stellar distribution is tilted relative to this hydrogen envelope. This is curious because the tilt does not seem likely to be a long-lasting feature in a galaxy that looks so well isolated. Perhaps KK~246 was disturbed by a relatively recent merger with another dwarf galaxy, though that would seem odd given the isolation. Perhaps, as Tinker and Conroy (2009) argued, the local void contains numerous dark matter halos that have too few stars and too little atomic hydrogen to be observable. Maybe KK~246 was disturbed by one of them. 

Rizzi, Tully, Shaya, et al. (2017) present an HST image of ZOA~J1952+1428; it too has the appearance of other low mass early-type galaxies. McIntyre et al. (2011) found that the mass of the atomic hydrogen envelope is $M_{\rm HI}=10^{7.0}M_\odot$. The optical luminosity, $L_B=10^{7.5}L_\odot$, suggests the mass in HI is less than the mass in stars. This is unusual; Bradford, Geha, and Blanton (2015) find that isolated low mass  galaxies typically have considerably more mass in atomic hydrogen than in stars.  One might wonder whether ZOA~J1952+1428 has been disturbed by an event that dissipated much of its HI, maybe supernovae, though that has not affected other dwarfs, or something external, though it appears to be isolated.

Another interesting object in the low density region toward the top of Figure~\ref{fig:LocalVol} is the spiral galaxy NGC~6946. It is marked by the open red circle at the largest positive value of SGZ. The ambient density is low there, but the image of this galaxy (to be seen on the web) looks much like the large spirals in the far more crowed region near the plane of the Local Supercluster that runs across the middle of the figure. A quantitative measure of that is the tight relation between the spiral galaxy circular velocity and baryonic mass (McGaugh 2020), which does not offer much room for sensitivity to environment. The atomic hydrogen surrounding NGC 6946 extends well beyond the starlight (Boomsma, Oosterloo, Fraternali, van der Hulst, and Sancisi, 2008), and it has the customary retinue of dwarf satellite galaxies (Karachentsev, Sharina, and Huchtmeier 2000). 

The galaxy NGC~6946 is a counterexample to one of those arguments that seem to make intuitive sense. In the standard cosmology the primeval departures from homogeneity are a random Gaussian process. For simplicity reduce this to two waves, a long wavelength one that represents ambient conditions, and a short wavelength one that represents the seeds of galaxy formation. Suppose a seed that happened to be near a maximum density in the long wavelength component, the ambient density, could produce a large galaxy like NGC~6946. That same seed that happened to be near a minimum of the ambient density would have a smaller total density; it would end up as a dwarf. The picture looks reasonable. It works for the most massive galaxies, which are found in particularly dense regions such as clusters of galaxies. It makes sense put another way: a galaxy might be expected to grow larger where the ambient density is larger and better able to supply matter to the growing galaxy. But this intuition does not account for the presence of NGC~6946 in the low density region above the Local Supercluster in Figure~\ref{fig:LocalVol}. And it does not account for the general evidence of similar space distributions of $L\sim L_\ast$ galaxies and the far more numerous dwarf galaxies (Davis, Huchra, Latham, and Tonry 1982;  Zehavi, Zheng, Weinberg, et al. 2011.) This failure of intuition is an anomaly.  

Conditions in the Local Void are different from our neighborhood, to judge by the scarcity of galaxies. What might be new and interesting there? Maybe dark matter halos with HI but no stars, or dark matter halos without baryons, or even HI clouds without dark matter? 

Arrays of telescopes such as MeerKAT are sensitive to the 21-cm line from atomic hydrogen, and  will be surveying the Local Void as part of scans of all the sky at all the radio frequencies accessible to the telescopes. But the Local Void is interesting enough to justify a Grand Project: a far deeper search for 21-cm sources confined to the part of the sky and the range of redshifts of the Local Void. This restricted use of an important facility would limit its production, but consider the compelling scientific interest in this unique opportunity to probe a void as deeply as possible. 

The image of the dwarf KK~246 in the Local Void is easy to see on the digitized ESO sky survey  plates (when I have been shown where to look), and I suppose there is not likely to be more dwarfs this luminous in the Local Void at this distance and not obscured by dust at low galactic latitude. The fainter Local Void dwarf ZOA~J1952+1428 was discovered as a 21-cm source, but the HST image in Rizzi et al. (2017) certainly looks like an unambiguous detection of the stars. This means an optical to infrared search for more of these faint dwarfs in the Local Void is technically feasible, if given expensive resources. 

The Local Void is particularly interesting because it can be examined in particular detail. Why are galaxies of stars so scarce in this void? Why does the spiral NGC~6946 with its retinue of dwarfs show so little indication of having been affected by its isolation? Why do the two dwarfs in the part of the Local Void pictured in Figure~\ref{fig:LocalVol} seem unusual despite their apparent isolation? What else is in this void?

\section{Galaxies}\label{sec:galaxies} 

A century of research on the nature of galaxies\footnote{A century ago \"Opik (1922) turned earlier thoughts that the spiral nebulae might be other galaxies of stars into a quantitative demonstration. \"Opik started with the assumption that the Andromeda Nebula M~31 ``consists of stellar matter similar to the matter of our Galaxy,'' with the same ratio of mass to luminosity as the estimate for the Milky Way. That combined with the angular size of M~31, its apparent magnitude, and the measured rotation velocity, from the Doppler shift, yields a useful estimate of the distance and mass of this galaxy. (Showing how this follows is a good exercise for the student.)  \"Opik's distance is half the correct value, an impressive advance at the time, and clear evidence that M~31 is a massive galaxy of stars, comparable to the size of the Milky Way.} has yielded a rich phenomenology and the challenge of understanding how or whether the phenomenology agrees with the cosmology. The complex nature of galaxies limits this test, but there are regularities that are useful hints to how the galaxies formed, which in turn offer guidance to whether the properties of galaxies fit what is expected in the standard $\Lambda$CDM cosmology. An example from the late 1990s is the prediction by Neta Bahcall and colleagues that in the Einstein-de Sitter model the masses of rich clusters of galaxies grow more rapidly than observed (Bahcall, Fan, and Cen 1997). This was credible early evidence that the mass density is less than predicted by the Einstein-de Sitter model that was popular then. The evidence remains credible and an example of how galaxies serve to test cosmology. Deciding which galaxy regularities, or curiosities, seem  worthy of closer attention must be a matter of taste, of course. I offer the following potentially informative lines of thought. 

\subsection{Early and Late Type Family Resemblances} \label{sec:earlyandlate}

Galaxies, like snowflakes, are all different if examined closely enough. You can see this by looking at the images of nearby $L\sim L_\ast$ galaxies to be found on the web. Among them are distinctly odd objects, but they look odd because they do not resemble the great majority of nearby galaxies that are readily classified as either elliptical or spiral, or in common usage early or late.\footnote{A century ago the two morphological classes, spiral and elliptical, were well known, as seen in Wolf's (1908) sketches showing examples that include the elliptical NGC 4494 and the spiral M~101. Jeans, Hubble, and others thought a galaxy might evolve from one type to the other, hence the names early and late. This now seems unlikely, but use of the names remains common.} In broad terms, another way to put it is that, with occasional exceptions, late type $L\sim L_\ast$ galaxies are gas-rich and early types are gas-poor. Stars are forming frequently enough in a gas-rich galaxy that the short-lived massive luminous blue stars tilt the spectrum of the galaxy to the blue; it is said to be in the blue cloud in a scatter plot of galaxy color and luminosity. A gas-poor galaxy has relatively few massive young blue stars; it fits in the red sequence in this color-luminosity plot (e.g., Salim, Rich, Charlot, et al. 2007, Fig. 1). S0 galaxies are a complication, but they are not common nearby, and they do not figure much in this essay (which might be a serious omission, as noted in footnote~\ref{fn:S0s}).

The galaxies in each of the two distinct types have their own family resemblances,\footnote{My use of the term, family resemblance, follows the Wikipedia interpretation of Ludwig Wittgenstein's thinking: family members share resemblances, or features, though no member need have all features. I take the term to be equivalent to family traits.} as in the red sequence and blue cloud. Tully and Fisher (1977) pointed out that the luminosity of a spiral galaxy is correlated with the circular velocity of the stars and gas in the disk. McGaugh (2020) presents the extension to include the mass in atomic hydrogen, which gives a tight correlation between the observed baryonic mass of a spiral and the circulation speed in its disk. This is a family resemblance, a characteristic of spiral galaxies. The analog for the early-type family began as the Faber and Jackson (1976) correlation between the luminosity and velocity dispersion of the stars in an elliptical galaxy. It was sharpened to the fundamental plane relating the elliptical galaxy luminosity, velocity dispersion, and radius (Dressler, Lynden-Bell, Burstein, et al. 1987; Djorgovski and Davis 1987). Bernardi, Nichol, Sheth, Miller, and Brinkmann (2006) show an example of this family trait, or regularity, and demonstrate that the regularity  is not sensitive to ambient density.

A notable family trait among ellipticals is the correlation of the spectrum of the galaxy with its stellar velocity dispersion: the greater the velocity dispersion the redder the mean spectrum (Zhu, Blanton, Moustakas 2010). But Zhu et al. show that at given velocity dispersion the mean spectra are very similar for ellipticals in more crowded and less crowded environments (apart from more  prominent H-$\alpha$ emission in field ellipticals). If ellipticals grew by dry mergers of star clusters that had a range of values of velocity dispersions, and hence a range of different spectra, then one might have predicted a sensitivity of the assembled elliptical to the present local situation, which might be expected to be correlated with the degree of merging. But the effect on the spectra is difficult to see.

The largest galaxies, with luminosities $L\sim 10L_\ast$ prefer dense regions. But the properties of the early and late-type galaxies with $L\sim L_\ast$, the ones that contribute most of the cosmic mean luminosity density, are insensitive to environment. This is notable evidence.

The early family type prefers denser regions. Early and late types have different life histories. Ellipticals have larger abundances of the alpha-process elements---carbon, oxygen, and so on---that are produced in early generations of massive stars, and lower abundances of the iron group elements that are more slowly produced in explosions of type~I supernovae. The abundance pattern in the spiral family brings to mind slower build-up of the elements, which agrees with the different distributions of stellar ages in the two families. 

\begin{figure}
\begin{center}
\includegraphics[angle=0,width=3.5in]{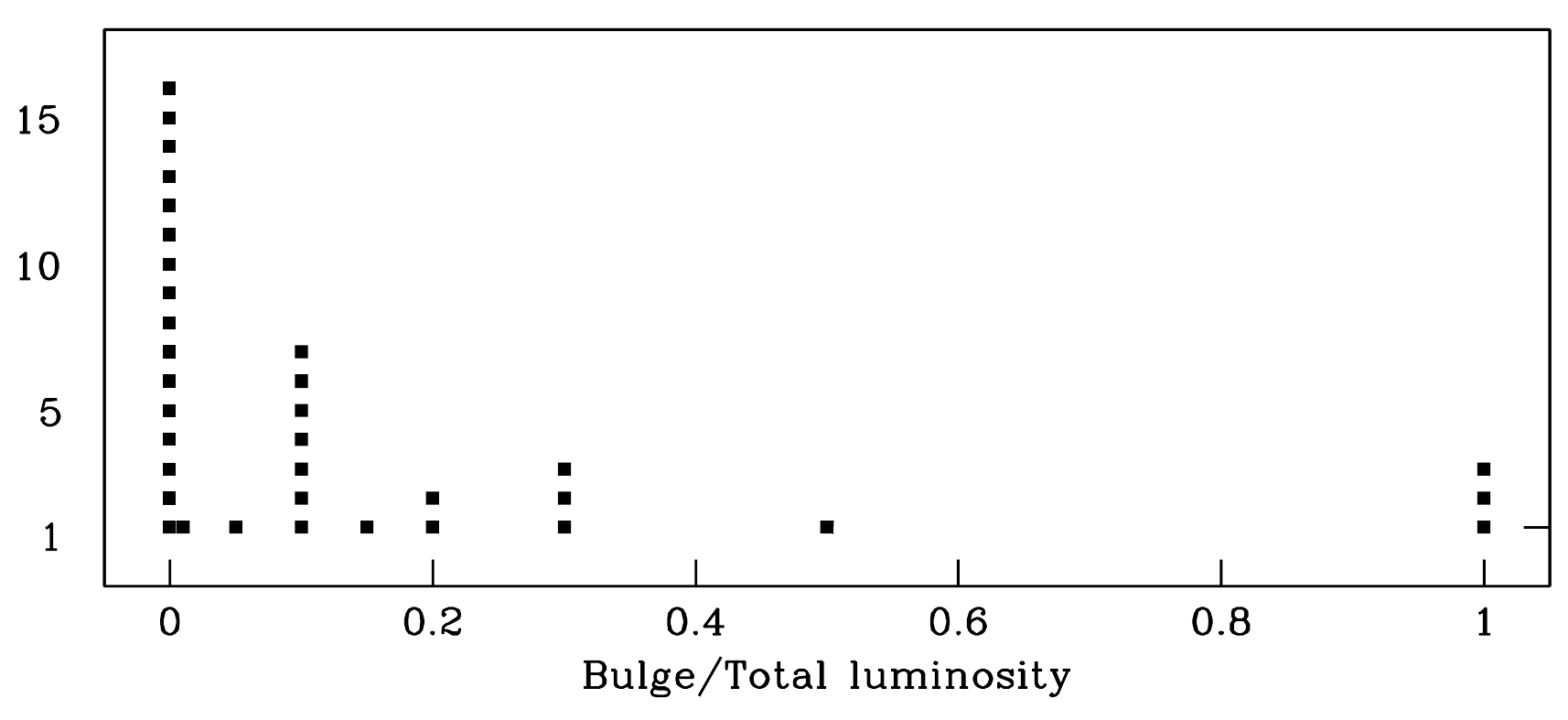}
\caption{\small Frequency distributions of ratios of bulge to total luminosities of nearby $L\sim L_\ast$ galaxies, from Peebles (2021).}\label{fig:BtoT}
\end{center}
\end{figure}

Figure~\ref{fig:BtoT} shows measured values of bulge to total luminosity $B/T$ for 32 galaxies that are within 10 Mpc distance and have luminosities $L_K > 10^{10}$ (from Kormendy, Drory, Bender, and Cornell 2010; and Fisher and Drory 2011). The sample is small, but beautiful images of the galaxies mentioned in the next paragraph are to be seen and admired on the web. 

The three ellipticals --- Centaurus~A, Maffei~1, and M~105 --- are in the modest peak at the right-hand side of the figure. The stars in these ellipticals are supported by near isotropic motions; we may say these stars are a hot component. The stars in the disk of a spiral galaxy are supported by rotation with a relatively small scatter around the mean: a cool component. The Sombrero Galaxy NGC~4594 at the center of the figure has $B/T=0.5$. It looks like a large spiral centered on an elliptical of similar size. Other disk galaxies further to the left in the figure, including the spirals M~31 and M~81, have a classical bulge, a hot component that rises above the disk. In these galaxies the bulge is more compact than in the Sombrero Galaxy. The pure disk spirals near the peak at the left-hand side of Figure~\ref{fig:BtoT} do not have an appreciable classical bulge, and their images look strikingly flat. Examples are M~101, NGC~253, and the edge-on galaxy NGC~4945. These pure disk spirals might have a pseudobulge, an unusually large surface brightness in the disk near the center. Authorities warn that observations of more distant galaxies at poorer spatial resolution may mistake a pseudobulge for a classical bulge; deciding which it is can be difficult. Some pure disk galaxies have a bar of stars that runs across the center of the galaxy; NGC~1300 is a pronounced example (at greater distance than the other galaxies mentioned here). 

There are exceptions to the two families. I mentioned the Sombrero Galaxy. The S0 galaxies have near featureless disk-like distributions of stars with bulges, 
giving the impression of spirals that have lost the spiral arms but kept remnants of the disk stars and the dust. Examples of nearby S0s are NGG~404, about 3~Mpc away, and NGC~2784, at about 10~Mpc. The S0 NGC~1460, at about 20 Mpc distance, is an elegant example of a barred galaxy without the spiral arms. These are rare exceptions to the population of galaxies outside clusters of galaxies; common in clusters. There are irregular galaxies; NGC~4490 looks like it is merging or falling apart, and the NASA/IPAC Extragalactic Database seems to be uncertain about the classification of the Circinus Galaxy. These exceptions are real but not common among nearby large galaxies. If the galaxies closer than 10~Mpc are a fair sample of the situation outside clusters of galaxies then they present us with clear and persuasive evidence that galaxies exhibit a distinct bimodality in their family traits.

It is said that elliptical galaxies formed by dry mergers, spirals by wet. Maybe an example is the situation in the group that contains the radio galaxy Centaurus~A. The two largest members are the elliptical NGC~5128, which is the radio source, and the spiral M~83.  Figure 1 of  Karachentsev, Sharina, Dolphin, et al. (2002) shows that most of the smaller galaxies around the late type M~83 also are late types, and most of the smaller galaxies around the early type Centaurus~A are early types. This agrees with the thought that early type galaxies grew by mergers of dry subhalos while late types grew by wet mergers. It is a description, of course, not an explanation.

People have been wondering about the origin of the early-late bimodality, and more broadly the Hubble sequence of galaxies, for the last century. Modern numerical simulations based on the $\Lambda$CDM cosmology capture aspects of the early and late morphologies (e.g., Vogelsberger, Genel, Springel, et al. 2014). Nelson, Pillepich, Springel, et al. (2018) show in their Figures~1 and~3 distributions of model galaxy color and stellar mass from their simulations of the formation of the central galaxies in dark matter halos. The distributions are quite similar to the Kauffmann, Heckman, White, et al. (2003) results from their analyses of the SDSS observations, which is encouraging. And their models at stellar masses $\sim 10^{10}M_\odot$ show bimodal morphologies. The empirical situation is richer, of course.  There are comparable numbers of the most massive galaxies with elliptical and spiral morphologies in the Huchra et al. 2012 catalog (as in Fig.~\ref{fig:morphologies} in Sec.~\ref{sec:LSC}, and in Ogle, Lanz, Appleton, Helou, and Mazzarella 2019), and there are clear examples of spirals and ellipticals at luminosities $L\lap L_\ast$. Understanding the distinct nature of galaxy bimodality remains an interesting challenge.

\subsection{What is the Separatrix for Bistable Galaxy Formation?}\label{separatrix} 

Current thinking, which is well motivated by the success of the standard cosmology, is that galaxies grew by gravity out of tiny primeval departures from an exactly homogeneous mass distribution, a stationary random Gaussian adiabatic process. The baryonic and dark matter gathered into mass concentrations, or halos, which grew more massive by merging with other halos and accretion of diffuse matter. Bayons settled, stars formed, and a protogalaxy grew into one or the other of the distinct galaxy families discussed in Section~\ref{sec:earlyandlate}.

Sidney van den Bergh's (1976) thinking about galaxy morphologies a half-century ago was that 
\begin{quotation}\noindent
\noindent Canonical views on galaxy evolution suggest that the present morphology of galaxies is predestined by the genetic heritage provided by initial mass and angular momentum. The results discussed above suggest that the evolution of galaxies is also substantially affected by environmental factors.
\end{quotation}
Both thoughts remain empirically well supported. We can add that, if galaxies grew by gravity out of small primeval Gaussian departures from homogeneity, then galaxy formation had to have been a bistable process. The point can be made a little more explicit by recalling the idea of bistable evolution and its separatrix in classical mechanics.

Suppose the state of a system is completely described by the values of $N$ components of particle positions and their $N$ canonical momenta. Let these $2N$ parameters be the coordinates in a $2N$ dimension phase space. The initial condition of the system is represented by its position in this space at a chosen starting time. Imagine an ensemble of initial conditions spread across phase space at this starting time. The equation of motion determines the evolution of the system, its path through phase space, from each initial position. In a bistable situation paths in phase space from the distribution of initial conditions arrive at one or the other of two (or more) basins of attraction. The separatrix is the boundary that separates initial positions in phase space that end up in one of the basins of attraction from the initial positions that end up in the other(s). The orbits of stars may have separatrices (e.g., Yavetz, Johnston, Pearson, Price-Whelan, and Weinberg 2021). The evolution of protogalaxies from their initial conditions is much more complicated; we must take account of dissipation, for example, and consider the complexities of stellar formation and its effects on the evolution of the galaxy. But the example from classical mechanics illustrates the concept of evolution of protogalaxies from initial conditions without manifest bimodality to a bimodal final state. It is what seems to have happened.

So in this way of thinking, what is the separatrix in galaxy formation that determines the evolution of a protogalaxy to a spiral or elliptical morphology? It cannot simply be the mass. There are spirals among the most luminous of galaxies, at $L \sim 10L_\ast$. (The distributions of the most massive spiral and elliptical galaxies relative to the plane Local Supercluster are shown in Fig.~\ref{fig:morphologies}.) At least some of these supermassive late types have the familiar two arms elegantly spiraling out from the center. An early example is UGC 2885 (Rubin, Ford, and Thonnard 1980); Ogle et al. (2019) catalogue others.  At lower stellar masses there are more spirals than ellipticals, but both types are observed. Van den Bergh (1976) had good reason to mention mass as part of the genetic heritage, but the story must be more complicated.

The disk of stars in a spiral galaxy is supported largely by rotation, while the stars in an elliptical are supported by a closer to isotropic distribution of orbits. Thus van den Bergh had good reason to consider that the separatrix is related to angular momentum. A dimensionless measure of the angular momentum $L$ of a galaxy is the combination $\Lambda = L~E^{1/2}G^{-1}M^{5/2}$, where $E$ the magnitude of the binding energy and $M$ is the mass (Peebles 1971). In analytic estimates and numerical simulations the distribution of $\Lambda$ is not bimodal (e.g., Efstathiou and Jones 1979). If, despite this, $\Lambda$ is the separatrix, the division between early and late types would require a sharp sensitivity to the values of $\Lambda$ and mass. A more likely picture along this line is that morphology is determined by ``the {\it coherent alignment} of the angular momentum of baryons that accrete over time to form a galaxy'' (Sales, Navarro, Theuns, et al. 2012). The investigation of galaxy morphology and halo spin in numerical simulations by Rodriguez-Gomez, Genel, Fall, et al. (2022) reveals a systematic difference of angular momentum of models classified as spirals and as ellipticals plus S0s. It is not yet bimodality in the spin-stellar mass plane (in their Fig.~1), but perhaps a step in this direction.

Environment matters. The giant $L\sim 10L_\ast$ galaxies in rich clusters likely formed by mergers of cluster members, meaning environment likely is the separatrix between these giants and ordinary $L\sim L_\ast$ galaxies. Maybe another example follows from the larger ratio of early to late types  in denser regions. For example, one might imagine that all protogalaxies began evolving toward the spiral morphology, but that violent mergers turned some proto-spirals into proto-early types. It would have happened more frequently in more crowded environments. But recall the separate family resemblances of spirals and ellipticals, which do not seem to be sensitive to environment. And consider the curious separation of early and late types in the Centaurus group (Sec.~\ref{sec:earlyandlate}).

The evidence is that the formation of galaxy morphologies was determined more by nature than nurture, and it is an interesting challenge to identify the character of the  separatrix. It might be some combination of mass, angular momentum, and environment, or maybe something completely different. The issue might be resolved by what is learned from numerical simulations of galaxy formation, or maybe by semi-analytic considerations of what appears to be happening. It is a fascinating opportunity for research, provided you bear in mind that people have been trying to solve the puzzle of the early-late bimodality for a long time. It means the resolution must be subtle, but surely it exists.

\subsection{Bulges and Disks of Spiral Galaxies}\label{sec:spirals}

 Numerical simulations of galaxy formation produce spiral galaxies that are impressively good approximations to what is observed, but there are three (or more; I invite suggestions) issues that are persistent enough to merit attention. One is that the central concentrations of starlight in simulated galaxies are overly luminous. Another is that the velocity dispersions of the stars moving in the planes of the disks of model spiral galaxies are unrealistically large. And a third is that we do not know the separatrix responsible for the distinct galaxy bimodality. I review these issues in Peebles (2020b); they are outlined and considered further here. 

Early attempts to simulate galaxy formation encountered the problem that a cloud of baryonic matter --- gas and plasma --- with the mass and radius typical of an $L\sim L_\ast$ galaxy readily dissipates energy and collapses almost freely. Observed spiral galaxies must have avoided this overcooling problem, and it must be avoided in models for otherwise they would have overly prominent classical bulges or stellar halos. The problem has been tamed by adjustments of the prescriptions for star formation and models for the effects of the stars on the distributions of baryonic and dark matter, but the evidence is that the problem in simulations of galaxy formation persists.

Figure~\ref{fig:BtoT} shows the distribution of the ratio of bulge to total luminosities of the nearby large galaxies. For some galaxies we can add to the hot component in the bulge the hot stellar halo that spreads out to greater distances away from the disk. Estimates of the median value of the luminosity fraction of the stars in the two hot components, bulge plus halo, in spiral galaxies are (from Peebles 2020b)
\beq
\hbox{simulations: }{B+H\over T}\sim 0.45,\quad \hbox{observations: }{B+H\over T}\sim 0.18. \label{eq:BHoverT}
\eeq
The rest of the total luminosity, $T$, is assigned to the disk. The median of the  observed fraction is from examinations of ten nearby galaxies by Merritt, van Dokkum, Abraham, and Zhang (2016) and Harmsen, Monachesi, Bell, et al. (2017). The hot  fraction in simulations is from reports by Grand, G{\'o}mez, Marinacci, et al. (2017) and Garrison-Kimmel, Hopkins, Wetzel, et al. (2018) of the results of two large research programs. The greater hot fraction in simulations agrees with my visual impression of images of real and model spirals. You are invited to check your impression. 

A second anomaly is the large dispersion of model disk stars in the direction of the plane of a simulated spiral galaxy. An illustration uses a simplified model for a spiral galaxy in which the stars move in the plane of a disk with a flat rotation curve, constant circular speed $v_c$. This is a reasonable approximation to many observed and model spiral galaxies. I refer to Peebles (2020b) for the details of the results of computation of stellar orbits. The orbit of a model star is characterized by a parameter, $\epsilon$, that is a measure of the orbital angular momentum, with $\epsilon = 1$ for a star moving in a circular orbit and $\epsilon = -1$ for a star in a circular orbit but moving in the opposite direction from the mean motion of the disk stars.  

Numerical solutions give the rms radial velocities at two choices of this circularity parameter: 
\begin{align}
\langle (dr/dt)^2\rangle^{1/2} &= 0.32 v_c\hbox{ for } \epsilon = 0.9; \nonumber\\
	&= 0.45 v_c \hbox{ for } \epsilon = 0.8. \label{eq:radialveldispn}
\end{align}
I have not found discussions of disc star velocity dispersions in model disk galaxies. My estimate is that in recent suites of numerical simulations (Grand, et al. 2017; Garrison-Kimmel et al. 2018) the most promising of the distributions in $\epsilon$ for a spiral galaxy have at least a quarter of the stars at $\epsilon<0.9$, which means that the radial velocity dispersions are greater than about a third of the circular velocity $v_c$ in a quarter of the stars. A majority of the stars in a promising model have $\epsilon<0.8$, with radial velocity dispersion greater than about half the circular velocity. Observations of the distribution and motions of the stars in our neighborhood of the Milky Way Galaxy (Anguiano, Majewski, Hayes, et al. 2020) indicate the radial velocity dispersion in the thin plus thick disk stars is about $\sigma_r=43$~km~s$^{-1}$ with $v_c\sim 240$~km~s$^{-1}$. The models look much hotter. 

The evidence reviewed here is that the model galaxies that emerge from modern simulations of cosmic evolution have unacceptably large populations of stars with large velocity dispersions present in stellar halos, classical bulges, and disks. This looks quite different from the impression of cool populations of stars and gas in the common  nearby $L\sim L_\ast$ pure disk galaxies (with the small fraction of the stellar mass in hot stars in the halo). I do not know whether this anomaly has resisted persistent attempts at remediation, or perhaps it has been put aside pending explorations of how to deal with the many other complexities in modeling galaxy formation. But since the art of simulating galaxy formation has a large literature, and the problem with hot star populations in simulated spiral galaxies remains, it ranks as a serious anomaly. Perhaps the situation will  be resolved by further advances in numerical simulations based on the $\Lambda$CDM theory. Or again, maybe something is missing.

\subsection{Merger Trees and the Cosmic Web}

The phrases, ``merger tree'' and variants, and ``cosmic web,'' often figure in discussions of how the galaxies formed. Aspects of both are worth considering here. 

Merging certainly happens; a clear example of a violent merger is the nearby Antennae Galaxies. Ostriker (1980) introduced considerations of what the remnant of the merger of two $L\sim L_\ast$ galaxies, as in this example, would look like when it had relaxed to a close to steady state. Let us only note that the remnant would have a luminous stellar bulge and halo made of pre-existing stars, which certainly is not seen in the nearby pure disk galaxies. And if the remnant of merging spirals looked like an elliptical it would have an unusual mix of chemical elements. If the local galaxies are close to a fair sample of the situation outside clusters then relaxed remnants of mergers of $L\sim L_\ast$ galaxies are not common, because pure disk galaxies are common, and I expect we would have heard of it if ellipticals with odd chemical abundances were common. 

A more modest example of merging is the pure disk edge-on galaxy NGC 5907 with its stellar stream that we may expect  eventually will add to the stellar halo of this galaxy (e.g., van Dokkum, Lokhorst, Danieli, et al. 2020). The ample evidence of tails and streams of stars around ellipticals and spirals is suggestive of close passages and mergers (e.g., van Dokkum 2005). And the stellar halo of the Milky Way is growing by tidal disruptions of dwarf galaxies (e.g., Belokurov, Zucker, Evans, et al. 2006)

The concept of galaxy formation as a hierarchical merging process grew out of several considerations. As just noted, galaxies do merge. The distribution of galaxies on scales from $\sim 0.1$~Mpc to $\sim 10$~Mpc is well approximated as a scale-invariant clustering hierarchy. It seems natural that the hierarchy also formed at smaller scales and was erased by merging to form galaxies (Davis, Groth, Peebles, 1977). And hierarchical growth of clustering is seen in numerical simulations of cosmic structure formation. But although the merger tree concept is well motivated by theory and observation it is in the spirit of empiricism to ask whether, absent simulations but given our knowledge of the phenomenology, people would have been led to the hierarchical assembly picture. 

Absent the guidance of merger trees people might have settled on the Eggen, Lynden-Bell, and Sandage (1962) account of the formation of the stellar halo and disk of the Milky Way spiral galaxy by a closer to monolithic collapse. The theory would include occasional mergers of galaxies, as observed. It would include formation of substructure in protogalaxies to account for satellites and the spreading of baryons across the disks of the Milky Way and other spiral galaxies, but substructure need not be to the extent of formation of identifiable halos that merge to form identifiable halos in a merger tree. The Eggen et al. picture is more in line with the observation that the distributions of heavy elements in halo stars do not resemble the distributions in stars in dwarf satellites (e.g., Tolstoy, Hill, and Tosi 2009, Fig. 10). Our stellar halo instead would have formed in the closer to monolithic collapse Eggen et al. envisoned, and would have been salted by stars from accreted dwarfs that produced the ``field of streams'' (Belokurov, et al. 2006).

How are we to interpret the description of the formation of early type galaxies by dry mergers and the late type by wet mergers, those rich in diffuse hydrogen? It calls to mind formation of morphology by nature rather than nurture: protogalaxies that have dry or wet natures. That could happen in a merger tree, but more simply in a monolithic collapse.

Cowie, Songaila, Hu, and Cohen (1996) pointed out that lower mass galaxies on average formed the bulk of their stars later. This downsizing effect does not naturally follow from a hierarchical merger tree in which less massive halos formed earlier. The discrepancy need not be serious because galaxy formation is complicated, but it does call to mind a picture similar to Eggen et al. (1962).

Madau and Dickinson (2014) concluded that ``galaxies formed the bulk (75\%) of their stellar mass at $z < 2$.'' The growth of the stellar mass of a pure disk galaxy in the manner described by Madau and Dickinson cannot have been by the merging of subhalos, or galaxies, that contained many stars, because the stars would have ended up in stellar bulges or halos, which are not prominent in these galaxies. The stars in pure disk galaxies had to have formed out of gas or plasma that had settled to the disk, as in cool streams (e.g. Kretschmer, Dekel, and Teyssier, 2022). This has the flavor of the Eggen, Lynden-Bell, and Sandage picture. 

The phrase, ``cosmic web,'' also often figures in discussions of galaxy formation.  Bond, Kofman, and Pogosyan (1996) introduced the concept as an evocative description of the distribution of dark matter in numerical simulations: large and small concentrations of dark matter are connected by filaments of dark matter in a pattern that calls to mind a web. It also resembles the observed galaxy distribution. Bond et al. pointed out that the filaments in simulations might be observable by detection of atomic hydrogen along the filaments. One might imagine that the filaments also are threaded by a magnetic field, and maybe even cosmic strings. 

Detection of HI streams would be particularly interesting because the filaments are expected to be present, connecting collapsing concentrations of dark matter, if the dark matter is adequately well approximated as a continuous fluid with no pressure or viscosity and the initial conditions are continuous. Dark matter consisting of black holes would be expected to form streams if the masses were small enough, and separate dark matter halos if the masses were large enough.

Instrument arrays capable of detecting 21-cm radiation from the hydrogen in  primeval filaments of matter connecting dark matter halos may be becoming available (e.g., Tudorache, Jarvis, Heywood, et al., 2022; Greene, Bezanson, Ouchi, et al., 2022). It will be interesting to see the nature of the H{\small I} distribution between mass concentrations, and the constraint that places on the black hole mass in a black hole model of the dark matter. 

\subsection{Massive Black Holes}\label{massiveblackholes}

The large luminosities and compact natures of quasars led to the thought that these objects are powered by the energy released by the collapse of matter onto black holes with masses of perhaps a million solar masses (Salpeter 1964; Lynden-Bell 1969). The clear evidence now is that large galaxies contain central compact objects with masses in the range of $10^5$ to $10^{10} M_\odot$. The objects in the centers of the elliptical M~87 and our Milky Way spiral galaxy certainly are compact (Event Horizon Telescope Collaboration et al., 2019; 2022). It makes a good case that these two objects, and the ones in other galaxies, are supermassive black holes of the kind predicted by Einstein's general theory of relativity.

It was natural to suppose that matter would settle to the center of a galaxy and perhaps accumulate to the point of relativistic gravitational collapse. But that picture is complicated by the extreme difference between the density characteristic of an $L\sim L_\ast$ galaxy, perhaps $\sim 10^{-24}$ g~cm$^{-3}$, and the density characteristic of a $10^{9} M_\odot$ black hole, $\sim c^6G^{-3}M^{-2}$, roughly $1$~g~cm$^{-3}$. Feeding the growth of a supermassive black hole by dissipative settling of diffuse baryonic matter certainly is conceivable, but one might instead expect that the settling would result in multiple fragmentation and the formation of star clusters, as in the formation of the first stars (e.g., Abel, Bryan, and Norman, 2002). Statistical relaxation of the star cluster would in time produce core collapse to a central black hole surrounded by a nuclear star cluster. But this is only one of the several current lines of thought that are reviewed by Greene, Strader, and Ho (2020), along with a discussion of how these black holes are detected. Issues that seem particularly relevant are discussed here.

A clue to the formation of these central black holes is the relation between the  black hole mass and properties of the galaxy such as the bulge or spheroid mass or stellar velocity dispersion (e.g., Magorrian, Tremaine, Richstone, et al., 1998). We need more than one relation, because some of the pure disk $L\sim L_\ast$ galaxies that are common nearby have massive central black holes with at most modest classical bulges. The familiar example is the Milky Way Galaxy with its central bar, little starlight in a classical bulge, and clear evidence of a central black hole with mass $4\times 10^6 M_\odot$. Another is the galaxy NGC~4945 (Gaspar, D{\'\i}az, Mast, et al. 2022 and references therein). This galaxy is seen nearly edge-on, it looks wonderfully flat, and there is little indication of a concentration of starlight in a classical stellar bulge rising out of the disk. (You can see the image of this galaxy at \url{https://apod.nasa.gov/apod/ap220226.html}.) The evidence is that this galaxy contains an active galactic nucleus operating around a black hole with mass comparable to the one in the center of our Milky Way galaxy.

If evidence accumulates that every $L\sim L_\ast$ galaxy has a central supermassive black hole it will invite the thought that galaxies formed around black holes, whether supermassive or collections of less massive black holes that seeded their formation (e.g., Silk and Rees 1998; Carr and Silk 2018). The thought is encouraged by the observations of quasars at redshifts $z\sim 7$. These quasars presumably depended on the presence of massive black holes, at a cosmic time when the stellar masses of galaxies were much smaller than now. 

In the Local Void (discussed in Sec.~\ref{sec:localvoid}) galaxy formation seems to have been  suppressed. The thought that galaxies formed around black holes suggests that the Local Void contains black holes, seed or supermassive, that are centered on galaxies that are unusually small for their black hole masses. Maybe closer examinations of the space distributions and redshifts of the stars in the two void dwarfs at the top and to the left in Figure~\ref{fig:LocalVol}  could determine whether they contain unusually massive black holes for such small galaxies. There are other void galaxies to examine, and there is a considerable area of sky and range of redshifts in the Local Void for deeper surveys for HI sources, star clusters, and maybe even gravitational lensing by primeval black holes without stars, any of which could be a valuable clue to the origin of massive black holes.  

Seth, van den Bosch, Mieske, et al. (2014) present evidence of an ultra-compact dwarf galaxy with an exceptionally massive central compact object, maybe a black hole. It may be near the large elliptical galaxy M~60. Maybe this dwarf is the remnant of a galaxy whose growth around a particularly massive primeval massive black hole was interrupted by tidal stripping. It is easy to invent such scenarios, sometimes difficult to test them, but perhaps an essential part of the search for the explanation of these objects.

Are there normal-looking $L\sim L_\ast$ galaxies that do not have a central massive black hole? The relatively nearby and face-on spiral galaxy M~101 has lanes of dust that spiral in toward the center, ending at a star cluster with mass $\sim 4\times 10^6M_\odot$ (Kormendy et al 2010). It is not yet known whether there is a massive central black hole inside the star cluster. A detection, maybe from the shapes of integrated Doppler-broadened stellar lines, would be interesting, and a seriously tight upper bound on the mass of a central black hole would be even more interesting. 

The normal-looking spiral galaxy M~33, the third largest galaxy in the Local Group, does not have a central black hole more massive than about $2\times 10^3M_\odot$ (Gebhardt, Lauer, Kormendy, et al., 2001; Merritt, Ferrarese, and Joseph, 2001). It is difficult to imagine how a merger of two galaxies could have driven both black holes out of the remnant while leaving the pure disk morphology of this galaxy. The more likely interpretation is that the formation of M~33 did not require a primeval black hole. It would mean that the massive black holes in other galaxies need not have served as seeds for galaxy formation, but instead grew together with the galaxies. 

In the coevolution picture the accumulation of mass in a growing central black hole would be by dissipative settling and merging at rates that might be expected to differ from galaxy to galaxy. This could be compared to the formation of stellar bars that run across the centers of spiral galaxies: some dominate the shape of the spiral, some are less conspicuous, and others are not noticeable. So it might be with the formation of supermassive black holes. The correlation of  mass with host galaxy properties argues against a broad scatter of massive black hole masses at given galaxy mass, but it will be helpful to see the scatter of central black hole masses as a function galaxy properties in larger samples. It might serve as a test of the coevolution picture. And it is to be noted that the coevolution picture does not seem promising for M~33, because there is no evidence of a central massive black hole.


The LIGO detection of gravitational waves from the merging of black holes with masses $\sim 66$ and $85 M_\odot$ (the event GW190521 reported by Abbott, Abbott, Abraham, et al. 2020) was unexpected because the masses are intermediate between the black holes produced by the relativistic collapse of stars and the supermassive black holes in the centers of galaxies. Maybe they are in line with the idea that black holes in this intermediate mass range were seeds for the formation of supermassive black holes. There is a long span of logarithmic time from the end of inflation, or whatever saved us from the spacetime  singularity of the standard cosmology, to the formation of the isotopes of hydrogen and helium. During this time cataclysmic events of some sort might have produced massive black holes or their seeds.  The favorite thought is directed to the disturbances to the mass distribution as the universe expanded and cooled through first-order cosmic phase transitions. These transitions might have been violent enough to have produced seed black holes, maybe with a broad range of masses set by the variety of cosmic first-order transitions in standard particle physics (e.g., Cappelluti, Hasinger, and Natarajan, 2022). Another thought is that supermassive black holes or their precursors  formed during cosmological inflation (e.g., Kallosh and Linde 2022). Yet another is that seed mass black holes formed by collapse of the first generation of gravitationally bound clouds of hydrogen and helium with mass $\sim 10^5M_\odot$ set by the baryon Jeans length (Silk and Rees 1998). 

Considerable research on the theory and observations of supermassive black holes in the centers of galaxies has not yet produced a convergence to the theory the origin of these objects. Their presence remains an anomaly.

\subsection{Why the Characteristic Galaxy Luminosity?}\label{sec:Lstar}

The frequency distribution of optical luminosities of galaxies has a characteristic value, $L_\ast$. There are far more galaxies with luminosities less than $L_\ast$, but the $L\sim L_\ast$ galaxies produce most of the cosmic mean optical luminosity density. The largest known galaxies have luminosities $L\sim 10 L_\ast$. That factor of ten is a curiously abrupt cutoff compared to the broad range of luminosities of galaxies that are less luminous than $L_\ast$. What accounts for the value of $L_\ast$ and the cutoff at greater luminosities?

The abrupt cutoff was anticipated by Schechter's (1976) functional form for the galaxy luminosity function with its exponential cutoff. It grew out of a Press and Schechter (1974) argument that ``contains no ad hoc information about an initial spectrum of long-wavelength density perturbations.'' The counterargument in Peebles (1974) agrees with the more recent demonstrations that the formation of cosmic structure is sensitive to the form of the spectrum of primeval departures from homogeneity. The Press and Schechter argument nevertheless produced an analytic form for the luminosity function that remains broadly useful.

In pure dark matter numerical simulations of the growth of cosmic structure in the standard cosmology the halo mass function is not as abruptly truncated at the high mass end as the galaxy luminosity function (e.g., Garrison, Eisenstein, Ferrer, et al., 2018, Fig. 7). That need not be an anomaly; we must bear in mind  the complexities of how mass was apportioned to dark matter halos and the baryons to stars. But is the existence of the characteristic luminosity $L_\ast$ and the sharp upper cutoff in galaxy luminosities an accidental result of these complexities, or might both be more readily understandable in an improved cosmology?

\subsection{Why MOND is Successful but Unpopular}\label{MOND} 

The rotation speed $v_c$ of the stars and gas in the outer parts of the disk of a spiral galaxy typically is close to independent of radius. The standard interpretation is that the spherically averaged mass density in the outer parts of the galaxy varies with the distance $r$ from the galaxy as $\rho\propto r^{-2}$, which translates to gravitational acceleration $g\propto r^{-1}\propto v_c^2/r$, satisfying the condition that the speed $v_c$ is independent of distance from the galaxy. This is the flat rotation curve observed in many spiral galaxies. The starlight density in the outer parts of a typical spiral galaxy falls off more rapidly then $r^{-2}$. The standard remedy is the postulate that the mass in the outer parts of a galaxy is the nonbaryonic dark matter of the $\Lambda$CDM theory, the dark matter halo.

Milgrom (1983) introduced an influential alternative: modified Newtonian gravity, or MOND. Instead of the hypothetical dark matter Milgrom proposed that the rotation curve is flat in the outer parts of a galaxy because at gravitational acceleration less than a characteristic value, $a_0$, Newton's law is modified to gravitational acceleration $g = \sqrt{GMa_0}/r$. In this limit, and assuming most of the mass is within the radius $r$, then the speed in a circular orbit is $v_c=(GMa_0)^{1/4}$. If the luminosity $L$ of the stars in the galaxy is proportional to the mass $M$ in baryons, then MOND predicts that the value of the circular velocity $v_c$ in the outer flat part of the galaxy rotation curve scales with the luminosity of the galaxy as the universal form $v_c\propto L^{1/4}$. This is close to the empirical Tully and Fisher (1977) relation. It is even closer to the power law relation $v\propto M^{1/4}$ observed when $M$ is the mass in interstellar atomic hydrogen added to the mass in stars (McGaugh 2020). 

MOND was proposed after the discovery of the Tully-Fisher relation, but it is reasonable to count the observed $v_c\propto M^{1/4}$ relation as a MOND prediction that passes a tight test. One may ask why this successful prediction receives so little community attention.

Let us note first that in the standard $\Lambda$CDM cosmology the observed $v_c\propto M^{1/4}$ relation does not challenge the theory.  It is instead is a property of galaxies, along with the other family traits. The challenge is to explore in this theory how these family resemblances among galaxies grew as the universe expanded. It will be a crisis for $\Lambda$CDM if an explanation for the observed traits cannot be found within the theory. We know far too little about how galaxies formed to hope for a judgement on this point any time soon. 

An argument for MOND is the considerable literature on how the properties of galaxies, and groups and clusters of galaxies, can be understood in this picture. It is reviewed by Diaferio and Angus (2016) and Banik and Zhao (2022). If in an alternative world all our present  phenomenology of cosmic structure were known but nothing was known about the evidence of a relativistic evolving universe, MOND likely would be a community favorite. But in our world there are two serious reasons for limited interest in MOND.

First, the well-tested $\Lambda$CDM theory with its cold dark dark matter offers a ready and promising framework for development of a theory of the formation of cosmic structure, galaxies and all. It has attracted the attention of active and productive research groups. I have argued for problems with the results, but they are details that have not discouraged the research groups and I am hoping might guide us to adjustments that improve $\Lambda$CDM.

Second, Milgrom's (1983) MOND does not offer a ready framework for development of a viable cosmology. The approach explored by Angus (2009) and Diaferio and Angus (2016) follows thoughts about alternative gravity physics by Bekenstein and Milgrom (1984) and Milgrom (2010). A starting  postulate is that there is dark matter, in the form of thermal sterile neutrinos with rest mass 11~eV. The dark matter density parameter is close to the standard model. Gravity physics in this model is enough like general relativity at redshifts $z \gap 1000$, and the warm dark matter is enough like the cold dark matter of the standard cosmology, that the acoustic oscillation pattern imprinted on the CMB is close to the standard prediction and the measurements. This is an important result. At low redshift the neutrino hot dark matter would have largely escaped galaxy potential wells, and gravity physics would have become enough like Milgrom's original MOND that galaxy rotation curves fit measurements without dark matter in halos around galaxies and with the weaker gravity of MOND. This is viable, within the notions of the gravity physics in this picture, though contrived. It is not demonstrated in a theory that allows computations of predictions. Diaferio and Angus (2016) conclude that
\begin{quotation}
It remains to be seen whether [in the adopted gravity physics] gravitational instability in a universe filled with baryonic matter and one species of sterile neutrino with 11 eV mass can form the observed cosmic structure at the correct pace.
\end{quotation}
The neutrino dark matter might be trapped in growing clusters of galaxies, which could be a helpful feature for the application of MOND to clusters. But Diaferio and Angus caution that 
\begin{quotation}
 the ability to explain the cluster mass discrepancy does not directly imply that MOND combined with 11-eV sterile neutrinos can form clusters in a cosmological context.
\end{quotation}
 
To my mind it is exceedingly unlikely that an alternative gravity physics and cosmology without cold dark matter, or something that acts much like it, can fit the array of tests passed by the standard general theory of relativity applied to the $\Lambda$CDM cosmology. But good science seeks to replace intuition with worked predictions of adequately specified theories. As Diaferio and Angus conclude, it remains to be seen whether this can be done for a generalization of Milgrom's MOND. 

\section{Summary Remarks}\label{SummaryRemarks}

To prevent misunderstandings I repeat the conclusion in Section~\ref{sec:tests}, that the empirical tests give excellent reason to expect that a more advanced physical cosmology will look much like the theoretical $\Lambda$CDM universe, because many well-checked tests show that the $\Lambda$CDM universe looks much our universe. But the great progress in cosmology and the other physical sciences has left anomalies, some of which have been troubling for a long time. 

A century ago Pauli and Jordan understood the vast difference between the vacuum mass density allowed by the relativistic theory and the density suggested by quantum physics (Sec.\ref{sec:Lambda}). Applications of quantum physics have grown far broader, giving compelling evidence that this physics is a broadly useful approximation to reality, but I am not aware of a significant advance in resolving the quantum energy density problem, apart from the anthropic argument (in Sec.~\ref{sec:Anthropic}). 

A century ago \"Opik (1922) turned earlier thoughts that the spiral nebulae might be other galaxies of stars into a quantitative demonstration. The progress from there to a promising physical basis for analyses of how the galaxies formed was slower than the development of quantum physics, but we have now a well-tested cosmology that might be expected to be an adequate basis for a secure understanding of these objects. The starting fundamental goals for a theory of the galaxies are to understand galaxy stellar masses, the spatial distributions of the stars in galaxies, and the stellar motions. Modern numerical simulations have made encouraging progress to this end, but there are anomalies. The evidence I have seen is that, despite careful attention to the many details required for the numerical simulations, disk star velocity dispersions in the planes of model spiral galaxies are unrealistically large (Sec.~\ref{sec:spirals}). The overcooling problem remains, resulting in unrealistically large classical bulges and stellar halos (Sec.~\ref{sec:spirals}). Many of the nearby $L\sim L_\ast$ are strikingly flat, unlike model spirals. 

A century ago Wolf (1908) reviewed evidence of what proves to be the bimodal natures of galaxies. The far richer evidence we have now (Sec.~\ref{sec:earlyandlate}) presents us with an interesting opportunity: identify the separatrix that determines whether a protogalaxy evolves into a spiral or an elliptical (Sec.~\ref{separatrix}). How are we to understand the cool motions of stars in the wonderfully thin galaxies seen nearby, so very different from the motions of stars in elliptical galaxies? The problem has been known for a century, aspects of the situation are the subjects of many papers, but identification of the separatrix remains an open challenge.

Less familiar anomalies tend to be less secure because less thought has been given to the theory and observation, and we can add that it is natural to give less thought to what is contrary to accepted thinking. We have an example in Section~\ref{sec:distributions}, on the large-scale distributions of astronomical objects. In the standard cosmology clusters of galaxies formed where the primeval upward mass density fluctuations were unusually large. Within available accuracy this agrees with numerical simulations of cosmic structure formation. But why would the primeval mass distribution in the $\Lambda$CDM universe be so arranged that upward mass fluctuations capable of evolving into the clusters that at distances less than about 85~Mpc are present only near the extended plane of the Local Supercluster (Fig.~\ref{fig:LSCf} in Sec.~\ref{sec:LSC})? Why are the most massive elliptical galaxies within 85~Mpc close to this plane, while comparably massive spirals are not noticeably correlated with the plane (Fig.~\ref{fig:morphologies})? The powerful radio sources present in some galaxies are thought to be associated with massive central black holes. Many large galaxies contain these black holes. So why are the powerful radio sources within 85~Mpc present in a select few of the large galaxies, those near this plane? Is there something special about these black holes? Tully and Shaver knew aspects of this situation thirty years ago (Sec.~\ref{sec:LSC}). To judge by the sparse citations to Shaver's key point these phenomena have not captured the general attention of the community. But the phenomena surely are real, and interesting, and not likely to have grown out of the Gaussian, adiabatic, and near scale-invariant initial conditions of the present standard cosmology. 

On larger scales, the applications of scaling relations to convert cluster luminosities to distances for estimates of peculiar velocities yield indications of unreasonable bulk flows relative to the rest frame defined by the CMB (Sec.~\ref{sec:bulkflows}). This disagrees with observations of the effect of the motion of intracluster plasma on the CMB. This kinetic SZ effect indicates a reasonably small cluster bulk flow. Maybe the situation is confused by subtle systematic errors, though the peculiar cluster velocity measurements  have been carefully checked. If the anomalous peculiar velocities are confirmed it means the luminosities of first-ranked cluster members, and the properties of the intracluster plasma, are not well constrained by other cluster properties that are not sensitive to distance, maybe even that cluster properties depend on parameters that are not in the standard cosmology. The thought is speculative, but recall that there are cosmic magnetic fields, and maybe cosmic strings, which might connect clusters and perhaps set hidden parameters.

On scales comparable to the Hubble length the heliocentric dipole anisotropies of quasars and radio galaxies are about in the direction expected from the kinematic effect of our motion relative to the CMB, but the dipole amplitude is unacceptably large. If this were because of a real unexpectedly large dipole anisotropy in the mass distribution on the scale of the Hubble length then the standard cosmology would predict an unacceptably large local peculiar velocity. But recall that at distances $\lap 85$~Mpc the curious distributions of radio galaxies, massive elliptical and spiral galaxies, and clusters of galaxies encourage the thought that the positions of radio galaxies, and so likely quasars, are only weakly related to the mass distribution traced by $L\sim L_\ast$ galaxies. So we have a mystery: what would have caused the anomalous large-scale distributions of massive ellipticals, quasars, radio galaxies and clusters of galaxies? The evidence is that all these objects contain massive black holes. Establishment of the theory of how these black holes formed might help. 

The standard $\Lambda$CDM theory passes demanding well-checked tests that reenforce the expectation that some if not all of the curious phenomena discussed here will prove to be results of systematic errors and/or statistical fluctuations. But consider one example, the distributions of the clusters of galaxies and the most luminous galaxies at distances less than about 85~Mpc. The $\Lambda$CDM theory offers a good account of the number density of mass concentrations similar to rich clusters. This adds to the argument for this theory and against the proposed anomalies. But does the distribution of these mass concentrations that grow into the clusters shown in  Figure~\ref{fig:LSCf} in Section~\ref{sec:LSC} look reasonable? It certainly looks real. Luminous radio galaxies and the most massive elliptical galaxies also tend to be close to this plane. They are related, but identified  and cataloged in different ways. The consistent case for alignment of these object with the extended plane of the Local Supercluster is convincing. The odd thing is that the most massive spirals, and the galaxies that are most luminous at $60\mu$, are not noticeably concentrated to the plane.This curious situation is difficult to reject and maybe suggestive of something interesting to be discovered.

 Let us not blame the messengers for the problems with the properties of galaxies reviewed in Section~\ref{sec:galaxies} and the distributions of galaxies discussed in Sections~\ref{sec:distributions} and \ref{sec:localvoid}; research groups are doing the best they can with the theory they have. It has not escaped community attention that the extreme simplicity of the dark sector of the standard $\Lambda$CDM cosmology seems unlikely to be better than a crude approximation to reality, maybe crude enough to be an impediment to progress in understanding cosmic structure.
 
\section{Acknowledgements}

I am grateful to colleagues for guidance to issues arising. They include Jean Brodie and Elaina Tolstoy for advice about dwarf galaxies; Antonaldo Diaferio and Garry Angus or advice on the generalization of MOND; Simon Driver and Samir Salim for education about galaxy morphologies; Mike Hudson, Tod Lauer, and Kostas Migkas for explanations of cluster bulk flow measurements; Manoj Kaplinghat for discussions of supermassive black holes; Roya Mohayaee and Subir Sarkar for discussions of the kinematic dipole; Dylan Nelson for discussions of simulations of galaxy formation; Patrick Ogle for instruction on supermassive spiral galaxies; and Xavier Prochaska for comments on observations of extragalactic magnetic fields. I am particuarly indebted to Michael Strauss for guidance to the phenomena and Neil Turok for guidance to the theory. Turok encouraged me to write this considerable revision of the draft in Peebles (2021).

\label{lastpage}
\end{document}